\documentclass{article}
\usepackage[english]{babel}
\usepackage[letterpaper,top=2cm,bottom=2cm,left=3cm,right=3cm,marginparwidth=1.75cm]{geometry}
\usepackage{amsmath,amssymb,amsfonts,amsthm,authblk,mathtools,empheq,cite}
\usepackage{graphicx}
\usepackage{multirow}
\usepackage[colorlinks=true, allcolors=blue]{hyperref}

\newcommand{\R}{\mathbb{R}}

\newcommand*\diff{\mathop{}\!\mathrm{d}}
\newtheorem*{remark*}{Remark}
\usepackage[capitalise]{cleveref}
\usepackage{hyperref}
\usepackage{algorithm}
\usepackage{booktabs}
\usepackage[noend]{algpseudocode}
\usepackage{tikz}
\usetikzlibrary{arrows.meta,positioning,fit,calc,backgrounds}

\title{From Individual Trajectories to Population Densities: Weak-Form Inference of Heterogeneous Growth Laws}
\author[1]{Rainey Lyons\footnote{rainey.lyons@colorado.edu}}
\author[1]{Vanja Dukic}
\author[1]{David M.~Bortz}

\affil[1]{Department of Applied Mathematics, University of Colorado, Boulder CO 80309-0526, USA}

\begin{document}

\maketitle

\begin{abstract}
Single-cell time-lapse microscopy now provides high-resolution size trajectories for thousands of individual cells, yet principled methods for simultaneously learning the functional form of growth laws and the distribution of cell-to-cell physiological variability remain limited.  
We present a data-driven framework based on the recent Weak-form Estimation of Nonlinear Dynamics (WENDy) algorithm to (i)~select the best supported growth law from a set of candidate laws, and (ii)~recover individual-level growth parameters via empirical Bayes shrinkage within a random-effects model.  
We demonstrate the approach on both synthetic individual size data incorporating commonly found growth laws and recorded real size trajectories, showing that the method accurately recovers both growth-law structure and parameter heterogeneity at reasonable noise levels.
We finally end with a discussion on how the results obtained by the method here can be upscaled to the population level. 
\end{abstract}

\section{Introduction}

Quantitative models of biological growth often begin with a time series of size measurements from one or more individuals. 
At the level of a single cell, organism, or cohort member, this time series data is commonly described by an ordinary differential equation (ODE) of the form
\[
\dot{x} = g(x;\theta),
\]
where $x(t)$ represents the size of an individual at time $t$ and $g$ is the growth rate function parameterized by a vector $\theta$. 
However, modern biological experiments, particularly in the context of cell assays, increasingly provide hundreds or thousands of individual trajectories, not just population averages.
These experiments have revealed that it is not enough to describe the growth of the population by a single averaged growth law.
For example individual heterogeneity in growth rates has been recognized as a mechanism capable of producing dispersion and multi-modality in evolving size distributions \cite{BanksFitzpatrick1991QuartApplMath}, which is a feature that cannot be captured by a single deterministic growth rate.
More broadly, trait variation within the population can change ecological dynamics through mechanisms such as nonlinear averaging, selection among individuals, and induced correlations between traits and demographic performance.
Consequently, populations with identical average parameter values but different levels of variability can develop different size distributions, altering prediction accuracy and mechanistic understanding of the dynamics.
These considerations motivate the explicit estimation of the parameter distributions rather than reliance on a single representative growth curve.

At the population scale, hyperbolic structured population models \cite{DiekmannGyllenbergMetz2020JMathBiol} provide a natural framework for describing the evolution of size distributions.
In these models, transport in a physiological structure variable (e.g., size for size-structured models) is driven by a growth law, while birth, death, division, and coagulation-fragmentation terms encode demographic or cellular events as source, sink, or boundary processes.
Traditionally, the modeling assumptions made here are that individuals follow the same structure-dependent dynamics with some works allowing for nonlocal dependence on the population density \cite{AcklehIto2005JDifferEqu,DullGwiazdaMarciniak-CzochraEtAl2021,GwiazdaLorenzMarciniak-Czochra2010JournalofDifferentialEquationsb}.
These models are well studied from an inverse problem perspective.
Multiple works have studied how to infer model ingredients of such equations from aggregate density measurements \cite{Ackleh1999MathematicalandComputerModelling,LyonsDukicBortz2025PLoSComputBiol,DoumicPerthameZubelli2009InverseProblems,PerthameZubelli2007InverseProbl,BourgeronDoumicEscobedo2014InverseProblems}.
The case of heterogeneous growth rates is a much more recent field of study. 
Recent forward-problem analyses \cite{DoumicRatTournus2026RSocOpenSci} studied how heterogeneity in the individual growth rates can affect the Malthusian fitness of a cell population under two different test cases. 
Whereas for the inverse problem, \cite{BanksFitzpatrick1991QuartApplMath} studied methods for recovering a growth rate distribution from density data using a cohort mixture decomposition and was able to approximate the growth-rate heterogeneity from density data.
This literature establishes that population densities can, in principle, contain information about hidden growth and division mechanisms. 
In contrast, single-individual trajectory data provide a more Lagrangian perspective: rather than inferring growth mechanisms only from evolving histograms, one may attempt to infer the growth law and the distribution of individual parameters directly from marked trajectories, and then propagate those estimates back to the population level.

From the statistical side, individual heterogeneity has been studied through a nonlinear mixed-effects framework.
Random-effect formulations are a powerful tool for size trajectory data as they separate population level parameters from individual-specific deviations \cite{LindstromBates1990Biometrics,pinheiro2000mixed}. 
Particularly, these techniques have seen use in organism growth, where empirical-Bayes and random effects versions of von Bertalanffy and Gompertz growth models have been used to estimate individual variation within a population \cite{VincenziJesensekCrivelli,VincenziCrivelliMunchEtAl2016EcologicalApplications}.
While these approaches provide a principled statistical interpretation of individual variability, they typically require that the growth law be specified in advance and that parameter estimation proceed by repeated nonlinear fitting or forward simulation.
When several candidate growth laws are plausible, this can become computationally expensive and statistically delicate.
Moreover, the empirical spread of fitted individual parameters need not equal biological heterogeneity as it may also contain uncertainty due to measurement noise, noninformative trajectories, numerical approximation, and practical non-identifiability.

Classical forward solver based methods of direct nonlinear trajectory fitting have additional limitations in the present setting.
Output error methods generally require repeated numerical solves of the candidate dynamics, careful treatment of initial conditions, and nonlinear optimization over each individual. 
These difficulties are amplified when one is interested in between individual heterogeneity and in comparing several plausible growth laws. 
Weak-form scientific machine learning (WSciML) provides an alternative route.
Sparse identification methods such as SINDy \cite{BruntonProctorKutz2016ProcNatlAcadSci} and PDE-FIND \cite{RudyBruntonProctorEtAl2017SciAdv} demonstrated that parsimonious governing equations can be learned from data via sparse regression over candidate libraries.
Weak-form extensions such as WSINDy \cite{MessengerBortz2021MultiscaleModelSimul,MessengerBortz2021JComputPhys} replace pointwise derivative estimation by integration against smooth test functions and use integration-by-parts to move derivatives from noisy data to known test functions, thereby improving robustness to measurement noise. 
For parameter estimation methods, the Weak-form Estimation of Nonlinear Dynamics (WENDy) method \cite{BortzMessengerDukic2023BullMathBiol} estimates parameters to a known model directly from the weak-form residuals without any forward simulations while simultaneously accounting for the errors-in-variables structure in the regression.
These features make WENDy particularly well suited to the problem at hand as each trajectory can be converted into a weak-form regression problem where candidate growth laws can be fit in parallel and parameter uncertainty can be estimated without relying on unstable numerical derivatives. 

Since their development, WSciML methods have been extended well beyond benchmark ODE and PDE discovery. 
WSINDy has been adapted to online and streaming identification of PDEs \cite{MessengerDallAneseBortz2022Online}, data-driven model predictive control \cite{LopezEtAl2026WSINDyMPC}, stochastic network dynamics \cite{TianEtAl2026Network}, mean-field equations from particle data \cite{MessengerBortz2022MeanField}, and interaction laws from individual cell trajectories \cite{MessengerEtAl2022Anisotropic}. 
WSINDy has also been applied to multiple biological and physical applications including the coarse-graining of nearly periodic Hamiltonian systems \cite{MessengerBurbyBortz2024Hamiltonian}, the study of information requirements for model recovery in magnetohydrodynamics \cite{VaseyEtAl2025MHD}, the discovery of interpretable atmospheric models \cite{MinorMessengerDukicEtAl2025JGeophysResMachLearnComput}, discovery of low-dimensional soliton dynamics from scattering data \cite{MinorEtAl2026Soliton}, inference of structured-population growth, mortality, and birth processes \cite{LyonsDukicBortz2025PLoSComputBiol}, and the recovery of mean-field movement equations from sparse ecological data \cite{MinorEtAl2026Insect}. 
In parallel, WENDy recast weak-form regression as an errors-in-variables parameter-inference problem and has since been extended to nonlinear-in-parameter models, statistical assessment of estimator bias and coverage, and computationally efficient practical-identifiability analysis for partially observed systems \cite{BortzMessengerDukic2023BullMathBiol,RummelMessengerBeckerEtAl2025arXiv250208881,Heitzman-BreenDukicBortz2026BullMathBiola}. 
Collectively, this literature demonstrates that WSciML methods provide more than a noise-robust substitute for numerical differentiation; they offer a general framework for system identification across biological, physical, and engineering systems.

In this work, we develop a WENDy-based framework for learning heterogeneous (in parameter) growth laws from individual size trajectories. 
We assume that individual growth laws share the same functional form $(g)$, but their biological parameters (inherent growth rate, maximum size) vary across the population. 
Given a collection of these individual level trajectories, we use WENDy to fit and compare a finite set of candidate growth laws and estimate the individual level growth parameters for the selected model. 
We then use an empirical-Bayes random-effects shrinkage correction to separate true heterogeneity from the uncertainty induced by noisy finite trajectories. 
This correction is essential for the second component of this manuscript: the upscaling of this individual level inference to a population level growth rate.

We take the time to clarify the main contributions of this work. 
First, we formulate growth law selection from individual trajectories as a weak-form model-comparison problem rather than as a collection of derivative estimates or output error methods.
Using the computational efficiency and embarrassingly parallelizable structure of WENDy, we estimate individual growth parameters and map the resulting linear-in-parameter coefficients to biologically interpretable quantities. 
Next, we incorporate an empirical-Bayes random-effects layer that corrects the noise inflated parameter variance and yields an estimate of the true population-level heterogeneity. 
Both of these layers are directly based on the uncertainty estimates from the WENDy method.
Finally, we discuss multiple upscaling methods for propagating this inference to the population scale.
Although the examples and problem here focus on size growth of individuals, the framework can be tuned to apply more broadly to any system in which individual trajectories are observed and population dynamics are strongly affected by heterogeneity.

The manuscript is organized as follows.
Section~\ref{sec:assumptions} formalizes the problem and introduces important notation and assumptions used throughout the work.
Section~\ref{sec:wendy_methods} describes in detail the methods used at the individual level including the WENDy method, model selection criteria, and mixed-effects correction for population level heterogeneity.
Section~\ref{sec:results} then showcases the results of the method and tests the results on both synthetic data (generated from commonly used growth laws) and real biological measurements.
Section~\ref{sec:population_level} then discusses methods to upscale the learned growth laws to the population density level.
Finally, Section~\ref{sec:discussion} provides a summary of the results, limitations of the methods, and an outlook of the open problems with the methods presented here.

\section{Problem statement, assumptions, and notation}\label{sec:assumptions}


We assume we are given a time-series measurement of (marked) individual sizes from an experiment of one species.  
These size measurements of individual $i = 1,2,\dots,I$  are denoted by $\{x_m^{(i)}\}_{m=1}^{M_i}$, where $x_m^{(i)}$ denotes the size (e.g., length, mass, volume) of individual $i$ at time $t_m^{(i)}$ with some fixed sampling rate $\Delta t$. 
We further assume that for this species there exists a parameterized, universal smooth growth law~$g$ such that
\begin{equation}\label{eq:growth_ode}
  \dot{x}^{(i)} = g\!\left(x^{(i)};\, {\theta}^{(i)}\right),
\end{equation}
where ${\theta}^{(i)}\in\R^{p}$ is a vector of individualistic parameters. 
That is, each individual's size evolves according to the same underlying dynamical system, but with parameters that vary across individuals which capture physiological differences such as inherent growth rate and maximal size.

Our goal is twofold: (i)~select the functional structure of $g$ from a family of candidate models $\mathcal{G}$, and (ii)~recover the distribution of the parameters, $\mathcal{P}$, across the population.  
Regarding this distribution, many possible assumptions could be made regarding the correlation between the daughter and mother's parameters.  
For this preliminary work, we take as a working assumption that the parameters for each individual are i.i.d. and leave this dependence as a study for future work with the ideas of following the approach from \cite{DoumicHoffmannKrellEtAl2015Bernoulli}.

In the sections to follow, we will use multiple, mathematically equivalent, forms of the growth model, $g$, in \eqref{eq:growth_ode}.
This is because the method of choice works best when the model is linear-in-parameters (LiP), but the candidate models may not always depend linearly on the biologically interpretable parameters.
To denote this change of form in a general way, we make use of the variables $\theta$ and $\mathbf{w}$ to represent the vector of biologically interpretable parameters and the vector of LiP parameters, respectively.
As an example, for the logistic equation $\dot{x} = rx(1 - x/S_{\max}) = w_1x + w_2x^2$, we have $\theta = (r,S_{\max})^\top$ and $\mathbf{w} = (w_1,w_2)^\top$ and we write $g(x;\theta)$ to mean the former equation and $g(x;\mathbf{w})$ to mean the latter.

\subsection{Overview of the method}\label{ssec:overview}
Before diving into detail in Section~\ref{sec:wendy_methods}, we provide a brief overview of the method used in this paper.
The proposed method will have two main components.
First, we aim to select a proper growth model from a class of candidate laws, $\mathcal{G}$.
This is done through utilizing the computational efficiency of the Weak-form Estimation of Nonlinear Dynamics (WENDy) method and the embarrassingly parallelizable nature of the problem.
Using WENDy to quickly fit a large number of trajectories, we can then select a model based on the summed per-trajectory weak-form BIC.

\begin{figure}[tbp]
  \centering
  \hyphenpenalty=10000\exhyphenpenalty=10000
  \scalebox{0.82}{%
  \begin{tikzpicture}[
    font=\small,
    >={Stealth[length=2.6mm,width=2.2mm]},
    bar/.style    ={rounded corners=2.5pt, draw, semithick, align=center,
                     inner sep=5pt, text width=116mm, fill=black!5, draw=black!55},
    img/.style    ={inner sep=0},
    t1/.style     ={align=center, text width=118mm, text=blue!45!black, font=\small},
    t2/.style     ={align=center, text width=118mm, text=orange!60!black, font=\small},
    arr/.style    ={->, thick, draw=black!70},
    grpMS/.style  ={rounded corners=4pt, draw=blue!50!black,   solid, line width=0.7pt, fill=blue!3},
    grpPE/.style  ={rounded corners=4pt, draw=orange!70!black, solid, line width=0.7pt, fill=orange!4}
  ]
 
  \node[bar] (inputs)
        {\textbf{Inputs:} individual size trajectories $\{x^{(i)}_k\}_{i=1}^{I}$ and candidate growth laws
         $\mathcal{G}$ (exponential, logistic, Gompertz, linear \& metabolic von Bertalanffy)};

  \node[t1, below=8mm of inputs] (h1)
        {\textbf{Model selection.}\ WENDy fits every trajectory under every $g_\ell\in\mathcal{G}$ in parallel;
         the summed per-trajectory weak-form BIC selects the growth law $\widehat{g}$.};
  \node[img, below=2.5mm of h1] (i1) {\includegraphics[width=120mm]{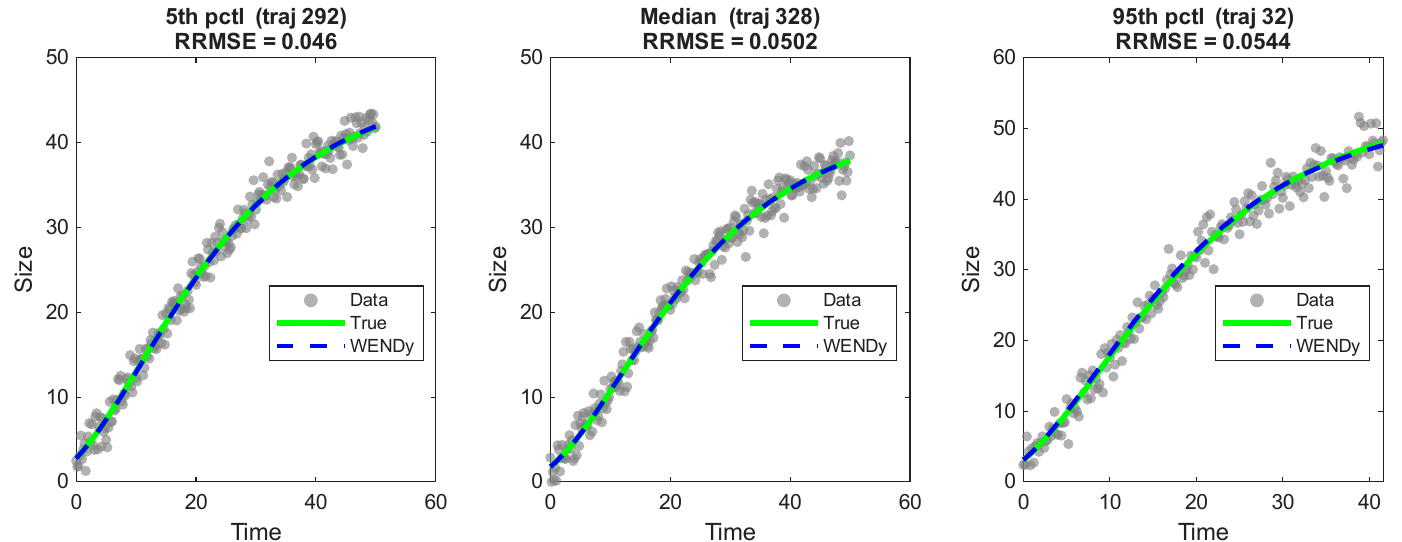}};
  \begin{scope}[on background layer]
    \node[grpMS, fit=(h1)(i1), inner sep=3.5mm] (b1) {};
  \end{scope}
 
  \node[t2, below=9mm of b1] (h2)
        {\textbf{Parameter estimation \& empirical-Bayes shrinkage.}\ Random effects separate true
         between-individual heterogeneity from noise, recovering the parameter distribution $\widehat{\mathcal P}$.};
  \node[img, below=2.5mm of h2] (i2) {\includegraphics[width=120mm]{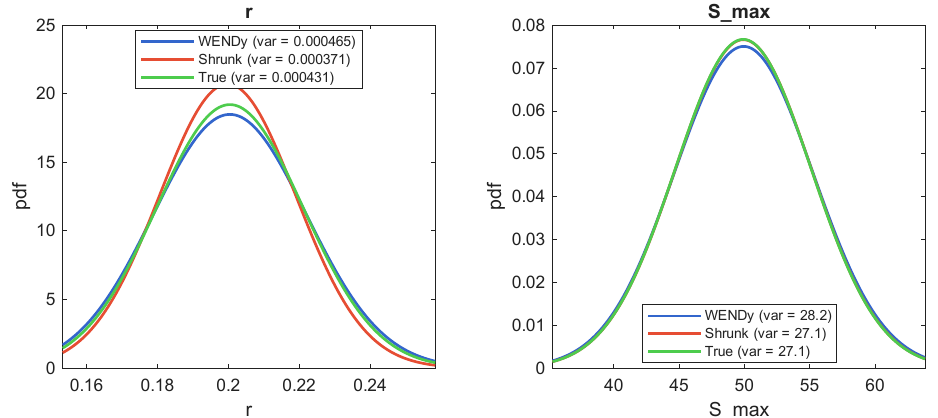}};
  \begin{scope}[on background layer]
    \node[grpPE, fit=(h2)(i2), inner sep=3.5mm] (b2) {};
  \end{scope}
 
  \node[bar, below=9mm of b2] (out)
        {\textbf{Outputs:}\ selected law $\widehat{g}$  $\bullet$  individual fits $\{\widehat{\mathbf w}^{(i)}\}$  $\bullet$ 
         heterogeneity $(\widehat\mu,\widehat\tau,\widehat{\mathcal P})$\ \ $\bullet$\ \ shrunk estimates $\widetilde\theta^{(i)}$  $\bullet$ 
         accuracy (RRMSE, $E_2$)};
 
  \draw[arr] (inputs) -- (b1);
  \draw[arr] (b1)     -- (b2);
  \draw[arr] (b2)     -- (out);
 
  \end{tikzpicture}%
  }
 
  \caption{Overview of the method. From marked individual size trajectories and a
  finite candidate growth-law class $\mathcal{G}$, the \emph{model-selection} stage
  fits every trajectory under every candidate with WENDy in parallel and selects the law
  $\widehat{g}$ by the summed per-trajectory weak-form BIC (top: data, true, and
  WENDy-fit trajectories at the 5th, 50th, and 95th RRMSE percentiles). The
  \emph{parameter-estimation} stage maps the fitted weights to biological parameters
  and applies an empirical-Bayes random-effects shrinkage that separates true
  between-individual heterogeneity from noise-induced spread (bottom: recovered $r$
  and $S_{\max}$ distributions for the raw WENDy, shrunken, and true parameters). The
  panels show a representative synthetic example (metabolic von Bertalanffy law, 5\%
  additive noise).}
  \label{fig:overview}
\end{figure}
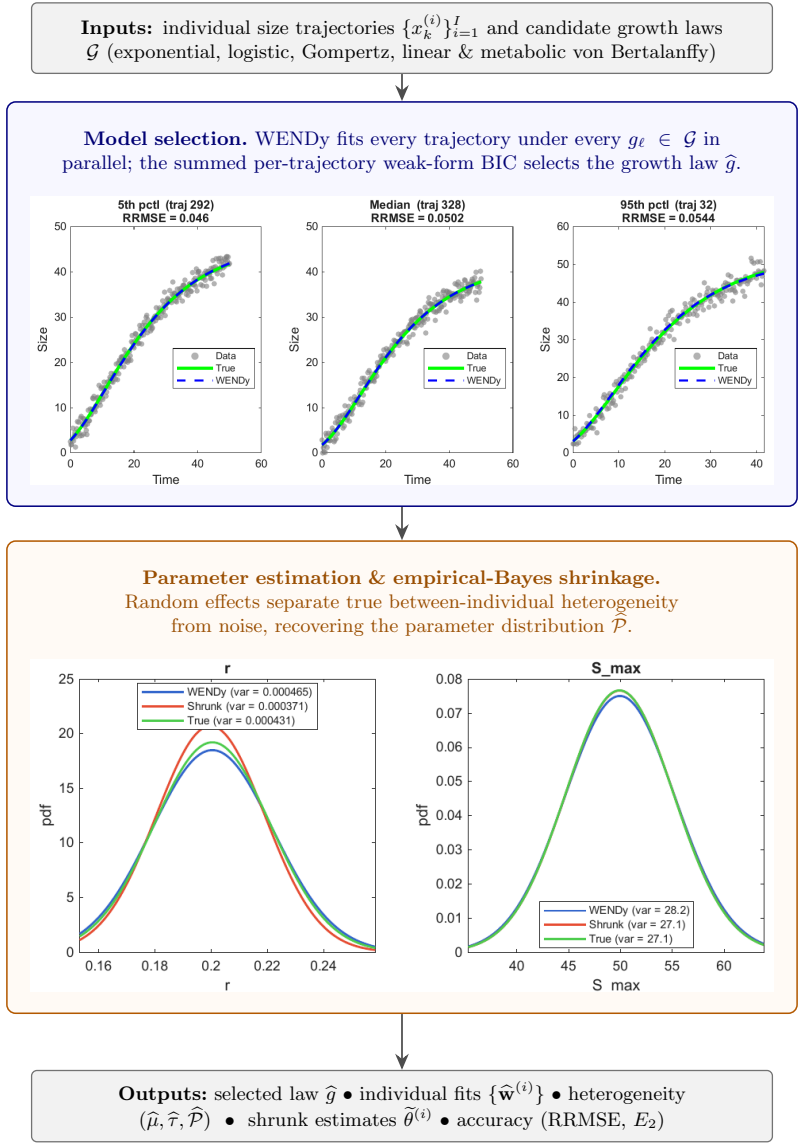

After a model has been selected, the learned WENDy parameters (after transformation) provide an estimate of the distribution of the biological parameter $\mathcal{P}$. 
However, as this estimate was done at the individual level, its variance can be corrupted by measurement error. 
Therefore, the second step of the proposed method is to correct the variance of the distribution via empirical Bayes shrinkage.
This combines the information provided by the individual level fits from WENDy with the information at the population level and allows us to correct the estimated variance.
A graphical overview of the method is provided in Figure~\ref{fig:overview}. 


\section{Methods}\label{sec:wendy_methods}

In this section, we provide a brief overview of the methods used in this manuscript.
We begin with a description of the WENDy algorithm for parameter estimation. 
We then discuss how the algorithm is used for the model selection step and then for recovering individual trajectory fits.
Finally, we discuss how these parameter fits are adjusted to account for the mixed effects due to measurement noise.
For detailed descriptions of the WENDy framework, we refer the reader to 
\cite{BortzMessengerDukic2023BullMathBiol,RummelMessengerBeckerEtAl2025arXiv250208881,MessengerDwyerDukic2024JRSocInterface}.

\subsection{Weak-form Estimation of Nonlinear Dynamics (WENDy)}\label{subsec:wendy}

Following the notation from before, let \(x^{(i)}(t)\) denote the size of individual \(i\) at time \(t \in (0,T)\), and assume
\begin{equation}\label{eq:wendy_model}
    \frac{\diff}{\diff t}x^{(i)}(t)
    =
    g\!\left(x^{(i)}(t);\mathbf{w}^{(i)}\right),
\end{equation}
where $g$ is written in a linear-in-parameters form, e.g., 
\(g(x) = \sum_{j=1}^Jw_j f_j(x).\)
Given noisy observations
\[
    \{x_m^{(i)}\}_{m=1}^{M_i}
    \approx
    \{x^{(i)}(t_m^{(i)})\}_{m=1}^{M_i},
\]
WENDy estimates \(\mathbf{w}^{(i)}\) without directly approximating \(\dot{x}^{(i)}\), avoiding known problems with numerical differentiation of noisy data.
This is done by using the eponymous weak-form of \eqref{eq:wendy_model}.
Let $\{\phi_k\}_{k=1}^K$ be a collection of smooth test functions compactly supported in the observation interval.
Multiplying \eqref{eq:wendy_model} by $\phi_k$, integrating in time, and applying integration by parts yields the weak-form equation
\begin{equation}\label{eq:wendy_weak_form}
    -\int_0^T x^{(i)}(t)\dot{\phi}_k(t)\,\diff{t}
    =
    \int_0^T g(x^{(i)}(t);\mathbf{w})\phi_k(t)\,\diff{t},
    \qquad k=1,\dots,K.
\end{equation}
Since the derivative has been transferred from the measured state $x^{(i)}$ to the known test function $\phi_k$, the quantity $\dot{\phi}_k$ can be evaluated analytically.
Thus, the weak formulation replaces unstable derivative approximation with integration against smooth test functions.

Discretizing the integrals in \eqref{eq:wendy_weak_form} using the observed data yields a finite-dimensional residual system
\begin{equation}\label{eq:wendy_residual}
    \mathbf{r}(\mathbf{w})
    :=
    \mathbf{G}\mathbf{w} - \mathbf{b},
\end{equation}
where $\mathbf{b}\in\mathbb{R}^K$ stacks the weak time-derivative terms $    b_k \approx -\int x^{(i)}(t)\dot{\phi}_k(t)\,\diff{t},$
and $\mathbf{G}\mathbf{w}\in\mathbb{R}^K$ stacks the weak model terms $    (G\mathbf{w})_k
    \approx
    \int g(x^{(i)}(t);\mathbf{w})\phi_k(t)\,\diff{t}.$
The basic weak-form parameter estimate is obtained by solving
\begin{equation}\label{eq:wendy_nlls}
    \mathbf{w}^{\star}
    =
    \arg\min_{\mathbf{w}}
    \|\mathbf{r}(\mathbf{w})\|_2^2.
\end{equation}
For models which can be written in a linear in parameters form, this reduces to a linear least-squares problem,
\begin{equation}\label{eq:wendy_linear_system}
    \mathbf{b} = \mathbf{G}\mathbf{w}.
\end{equation}

In practice, the integrals are approximated using quadrature.
Let
\[
    \mathbf{x}^{(i)}
    :=
    \begin{bmatrix}
        x^{(i)}_1\\
        \vdots\\
        x^{(i)}_M
    \end{bmatrix}
    \in\mathbb{R}^M
\]
denote the observed time series, and let $\boldsymbol{\Phi}$ and $\dot{\boldsymbol{\Phi}}$ denote the matrices whose rows contain the test functions and their derivatives evaluated at the measurement times, scaled by the chosen quadrature weights (in our case, the trapezoidal quadrature scaling $q = (\frac12,1,1,\dots,1,1,\frac12)$).
For a model that is linear in parameters, the discrete weak-form system simply takes the form
\begin{equation}\label{eq:wendy_discrete_system}
    \mathbf{b}
    =
    -\dot{\boldsymbol{\Phi}}\mathbf{x}^{(i)},
    \qquad
    \mathbf{G}
    =
    \boldsymbol{\Phi}\boldsymbol{\Theta}(\mathbf{x}^{(i)}),
\end{equation}
where $\boldsymbol{\Theta}(\mathbf{u})$ is the feature matrix obtained by evaluating the prescribed model terms at the data.
For example, if $f_1(u)=u$ and $f_2(u)=u^{2}$ for the logistic equation, then
\[
    \boldsymbol{\Theta}(\mathbf{x^{(i)}})
    =
    \begin{bmatrix}
        f_1(x^{(i)}_1) & f_2(x^{(i)}_1)\\
        \vdots & \vdots\\
        f_1(x^{(i)}_M) & f_2(x^{(i)}_M)
    \end{bmatrix}.
\]

In this work, we use compactly supported piecewise polynomial test functions of the form
\[
\phi_k(t)
=
\begin{cases}
    C\left(1-\left(\frac{t-t_k}{r}\right)^2\right)^p,
    & |t-t_k|\leq r,\\
    0,
    & \text{otherwise},
\end{cases}
\]
where $t_k$ is the center of the test function, $C$ is a normalization constant, $p$ determines the smoothness of the test function, and $r>0$ is the radius of its support.
The support radius controls the balance between noise averaging and temporal localization: larger supports provide greater denoising through integration, whereas smaller supports provide more localized constraints but are more sensitive to measurement noise.
This compact structure is also important for mediating the quadrature error collected during the integral approximations above. 
Indeed, with  Euler--Maclaurin arguments (see \cite{MessengerBortz2021MultiscaleModelSimul}), the trapezoidal rule's quadrature error is determined by the smoothness of the test function and can therefore be mitigated by choosing a sufficiently smooth test function.
For the work here, we use the canonical method for radius selection presented in \cite{BortzMessengerDukic2023BullMathBiol}, however, we note that there are more methods for test function selection like the one presented in \cite{TranBortz2025arXiv250703206b} which may be used.

\subsection{Noise model and weighted WENDy estimation}\label{subsec:wendy_noise}

A central feature of WENDy is that it accounts for the fact that measurement noise enters both sides of the weak-form regression system.
Indeed, the vector $\mathbf{b}$ depends on the noisy state measurements, and the matrix or nonlinear map defining $\mathbf{G}$ also depends on the same noisy data.
Consequently, the weak-form regression problem is an \textit{errors-in-variables} problem, and ordinary least squares can be biased when the noise level is non-negligible.

We assume the observed data satisfies
\[
    \mathbf{x}^{(i)} = {\mathbf{x}^{(i)}}^{\star} + \boldsymbol{\varepsilon},
    \qquad
    \mathbb{E}[\boldsymbol{\varepsilon}] = 0,
    \qquad
    \mathrm{Cov}(\boldsymbol{\varepsilon}) = \Sigma_{x^{(i)}},
\]
where ${\mathbf{x}^{(i)}}^{\star}$ denotes the unknown noise-free signal.
In our artificial experiments below, we assume the true measurements are corrupted by additive Gaussian noise, where $\Sigma_x^{(i)} = \hat{\sigma}^2 I$ and $\hat{\sigma}$ is chosen such that the resulting noise ratio is of a specified level, $\sigma_{\mathrm{nr}}$, i.e., $\hat{\sigma} = \sigma_{\mathrm{nr}} \|{\mathbf{x}^{(i)}}^\star\|_{\mathrm{RMS}}$, see \cite{MessengerBortz2021MultiscaleModelSimul} for details.

Since the weak residual $\mathbf{r}(\mathbf{x}^{(i)},\mathbf{w})$
is a function of the noisy measurements, a first-order expansion around the noise-free trajectory gives
\[
    \mathbf{r}(\mathbf{x}^{(i)},\mathbf{w})
    \approx
    \mathbf{r}({\mathbf{x}^{(i)}}^{\star},\mathbf{w})
    +
    J_x^{(i)}(\mathbf{w})\boldsymbol{\varepsilon},
\]
where $J_x^{(i)}(\mathbf{w})$ is the Jacobian of the weak residual with respect to the data.
Therefore, the covariance of the residual can be approximated by
\begin{equation}\label{eq:wendy_residual_covariance}
    R(\mathbf{w})
    :=
    J_x^{(i)}(\mathbf{w})\Sigma_x^{(i)} J_x^{(i)}(\mathbf{w})^\top.
\end{equation}
Following the WENDy framework, this covariance model may be augmented by higher-order correction terms that account for nonlinear dependence of the weak residual on the noisy measurements.

Rather than solving the unweighted least-squares problem \eqref{eq:wendy_nlls}, WENDy approximately solves the generalized least-squares problem
\begin{equation}\label{eq:wendy_gls}
   \mathbf{w}^{\star}
    =
    \arg\min_{\mathbf{w}}
    \mathbf{r}(\mathbf{w})^\top
    R(\mathbf{w})^{-1}
    \mathbf{r}(\mathbf{w}),
\end{equation}
where $R(\mathbf{w})$ denotes the working covariance model for the weak residuals.
Because $R(\mathbf{w})$ depends on the unknown parameters, WENDy uses an iteratively reweighted least-squares procedure.
In other words, at each iteration, the residual covariance is estimated using the current parameter iterate, the weak-form system is pre-whitened using this covariance estimate, and the resulting weighted least-squares problem is solved to update the parameter estimate.
This procedure is repeated until convergence.

\subsection{Candidate growth laws and WENDy-based model selection}\label{subsec:wendy_model_comparison}

We can use WENDy to compare a finite collection of prescribed candidate growth laws.
Rather than learning an unrestricted governing equation from a large candidate library as with WSINDy\footnote{In the case all of the desired models are polynomial in structure such as the exponential, logistic, and linear von Bertalanffy growth laws, WSINDy is a perfectly applicable equation learning method. However, many of WSINDy's convergent properties fail when library terms are not monomial or trigonometric \cite{MessengerBortz2024IMAJNumerAnal}.}, we assume that the relevant growth law belongs to a finite model class $    \mathcal{G}    =    \left\{    g_1,\dots,g_L    \right\},$
where each model \(g_\ell\) corresponds to a prescribed functional form
\begin{equation}\label{eq:model_selection_candidate_law}
    \frac{\diff}{\diff t}x^{(i)}(t)
    =
    g_\ell\!\left(x^{(i)}(t);\mathbf{w}^{(i)}_\ell\right).
\end{equation}
Here, \(\mathbf{w}^{(i)}_\ell\in\mathbb{R}^{d_\ell}\) denotes the individual-specific parameter vector for model \(g_\ell\), and \(d_\ell\) is the number of fitted \emph{linear} parameters in that candidate law.
In other words,
\begin{equation}\label{eq:model_selection_reduced_library}
    g_\ell(x;\mathbf{w}^{(i)}_\ell)
    =
    \sum_{j\in S_\ell}
    w_{\ell,j}^{(i)} f_j(x),
\end{equation}
where \(S_\ell\) is the subset of basis functions associated with model \(g_\ell\) (see Table~\ref{tab:growthModels}).
When biologically appropriate, sign constraints are imposed during the WENDy regression so that the fitted coefficients correspond to admissible growth laws.

For each candidate model \(g_\ell\), we construct the reduced weak-form library associated with \(S_\ell\) and apply WENDy independently to each individual trajectory.
That is, for each pair \((i,\ell)\), WENDy produces an estimate
\[
    \widehat{\mathbf{w}}_\ell^{(i)}
    =
    \arg\min_{\mathbf{w}_\ell^{(i)}}
    \left\|
    \mathbf{r}_\ell^{(i)}(\mathbf{w}_\ell^{(i)})
    \right\|^2,
\]
with the weak residual
\begin{equation}\label{eq:individual_weak_residual}
    \mathbf{r}_\ell^{(i)}(\mathbf{w}_{\ell}^{(i)})
    =
    \mathbf{b}^{(i)}
    -
    \mathbf{G}_\ell^{(i)}
    \mathbf{w}_\ell^{(i)}.
\end{equation}
Here, \(\mathbf{b}^{(i)}\) contains the weak time-derivative terms for individual \(i\), and \(\mathbf{G}_\ell^{(i)}\) is the weak-form design matrix obtained from the active terms in model \(g_\ell\) as described in \cref{subsec:wendy}.

After fitting all individuals under model \(g_\ell\), we compute the individual weak residual sum of squares
\[
    \mathrm{RSS}_\ell^{(i)}
    =
    \left\|
    \mathbf{b}^{(i)}
    -
    \mathbf{G}_\ell^{(i)}
    \widehat{\mathbf{w}}_\ell^{(i)}
    \right\|_2^2,
\]
and define $ K_i := \dim\!\left(\mathbf{b}^{(i)}\right)$
to be the number of weak-form equations (or equivalently the number of test functions) used for individual $i$.
The aggregate weak residual sum of squares for model \(g_\ell\) is then
\begin{equation}\label{eq:aggregate_rss}
    \mathrm{RSS}_\ell
    =
    \sum_{i\in\mathcal{I}_\ell}
    \mathrm{RSS}_\ell^{(i)},
\end{equation}
where \(\mathcal{I}_\ell\) denotes the set of trajectories successfully fit under model \(g_\ell\).

As the parameters are estimated separately for each individual, a natural model-selection criterion would be the summed per-trajectory weak-form BIC,
\begin{equation}\label{eq:individual_bic}    \mathrm{BIC}_\ell^{(i)}
    =
    K_i
    \log\!\left(
        \frac{\mathrm{RSS}_\ell^{(i)}}{K_i}
    \right)
    +
    d_\ell \log(K_i),
\end{equation}
with aggregate score
\begin{equation}\label{eq:sumtraj_bic}
    \mathrm{BIC}_{\ell}^{\mathrm{sum}}
    =
    \sum_{i\in\mathcal{I}_\ell}
    \mathrm{BIC}_\ell^{(i)}.
\end{equation}
This criterion treats each individual trajectory as its own fixed-effects weak-form regression problem and then sums the resulting model-comparison scores across the population. 
However, as the test functions are allowed to overlap, neighboring rows of the weak-form system integrate largely the same measurements and are therefore strongly correlated. 
Instead, we replace $K_i$ in \eqref{eq:individual_bic} and \eqref{eq:sumtraj_bic} with the effective number of weak-form equations defined as 
\[
K^{(i)}_{\mathrm{eff},\ell} := \frac{K^2_i}{\|R^{(i)}_\ell\|^2_F},
\]
where $R^{(i)}_\ell$ is the residual covariance matrix given in \eqref{eq:wendy_residual_covariance} which is computed by WENDy and $\|\cdot\|_F$ denotes the Frobenius norm.

After model selection, the corresponding WENDy estimates
\(
    \{
    \widehat{\mathbf{w}}_{\widehat{\ell}}^{(i)}
    \}_{i\in\mathcal{I}_{\widehat{\ell}}}
\)
are retained as the individual-level growth parameters.
When relevant, these coefficient estimates are subsequently mapped to biologically interpretable parameters, such as an intrinsic growth rate $r^{(i)}$ and an asymptotic size $S_{\max}^{(i)}$.

\subsection{Estimating heterogeneity in the population}
\label{subsec:population_heterogeneity}

\begin{figure}[h]
    \centering
    \includegraphics[width=0.5\linewidth]{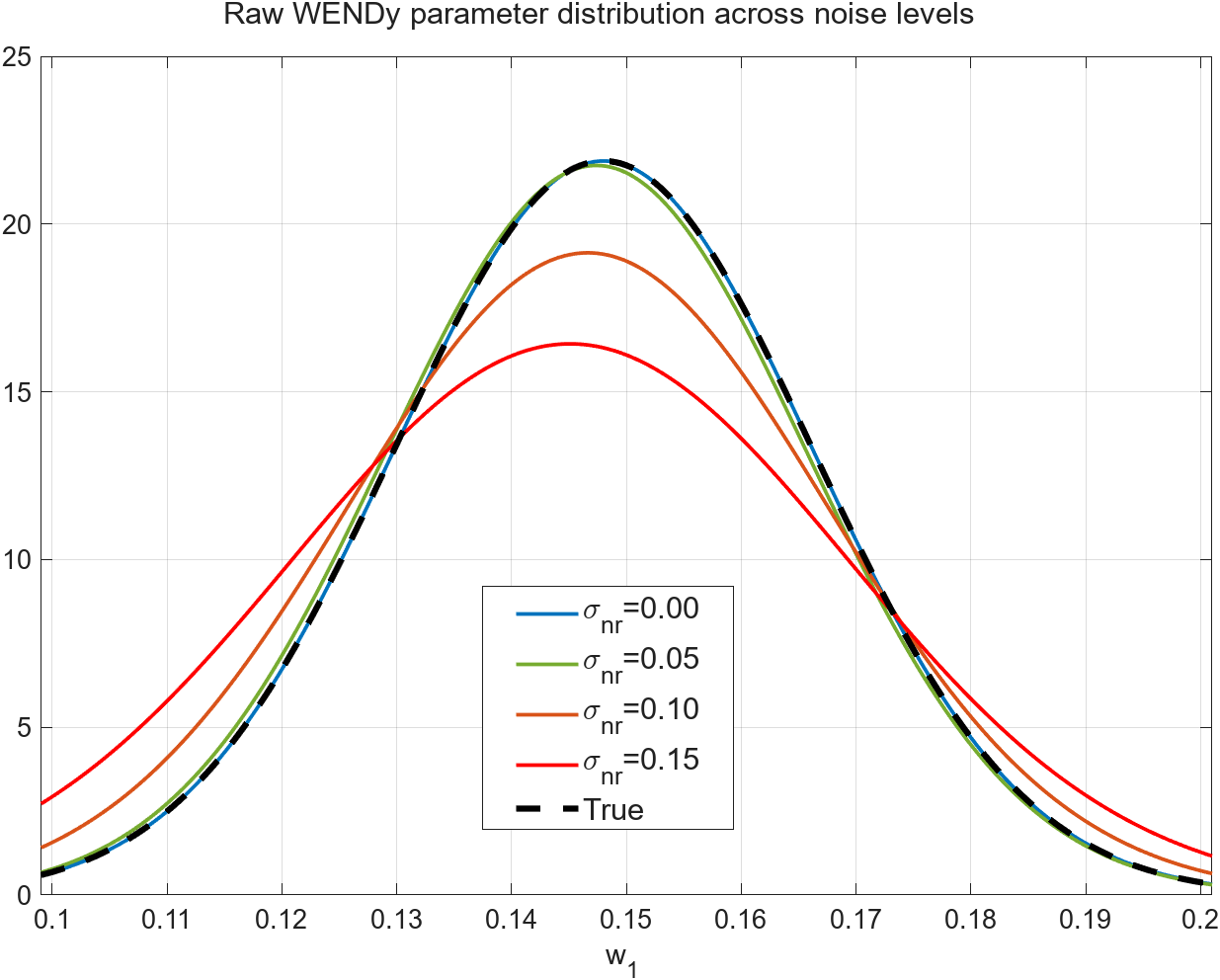}
    \caption{Learned parameter distribution for artificial exponential growth data from the raw WENDy algorithm presented in Section~\ref{sec:wendy_methods} as the measurement noise ratio $\sigma_{\mathrm{nr}}$ increases. Note for this model $\mathbf{w} = \mathbf{\theta}$.}    \label{fig:LearnedParamDist_differentNoise}
\end{figure}

After selecting a growth law, WENDy provides an individual-level coefficient estimate $ \widehat{\mathbf w}^{(i)}$
for each successfully fit trajectory $i\in\mathcal I_{\widehat \ell}$. These estimates are retained as the raw individual-level dynamical fits. However, the empirical distribution of the raw WENDy estimates generally overstates the true biological heterogeneity.
This is due to each estimate containing both true individual-to-individual variation and uncertainty induced by measurement noise, finite sampling, and numerical quadrature. 
This is demonstrated for the exponential equation in \cref{fig:LearnedParamDist_differentNoise} where we see the variance of the WENDy estimates increasing with the noise level.
Following similar ideas to those presented in \cite{VincenziJesensekCrivelli,VincenziCrivelliMunchEtAl2016EcologicalApplications,BenzekryLamontBeheshtiEtAl2014PLoSComputBiol}, we use an empirical-Bayes random-effects correction to estimate the population-level distribution of biologically interpretable growth parameters.

Let
\[
    \boldsymbol\theta^{(i)}
    =
    h_{\widehat\ell}\!\left(\mathbf w^{(i)}\right)
\]
denote the biological parameter vector associated with the selected law, where $h_{\widehat\ell}$ is a function which maps the linear WENDy weights $\mathbf{w}$ to the biological parameters of interest for model $\widehat\ell$. 
In the examples considered here, $\boldsymbol\theta^{(i)}$ consists of the intrinsic growth rate $r^{(i)}$ and, when the selected law has a finite asymptote, the asymptotic size $S_{\max}^{(i)}$. 
For example, for the logistic law, $    \dot x = w_1 x + w_2 x^2,$
the biological parameters are
\[
    r = w_1,
    \quad
    S_{\max} = -\frac{w_1}{w_2}, \quad
\text{and}\quad
h(\mathbf{w}) = \left(\begin{matrix}
    w_1\\
    -\frac{w_1}{w_2}
\end{matrix}\right).\]
We denote the raw WENDy estimate of these parameters by $\widehat{\boldsymbol\theta}^{(i)}
    =
    h_{\widehat\ell}\!\left(\widehat{\mathbf w}^{(i)}\right).$

The working assumption in this paper is that the biological parameters vary across individuals according to a population-level distribution. 
In the implementation used here, we assume that each biological parameter is normally distributed across the population. 
That is, for a scalar biological parameter $\theta$ such as $r$ or $S_{\max}$, we assume 
\begin{equation}
    \theta^{(i)} \sim \mathcal N(\mu,\tau^2),
    \qquad
    i\in\mathcal I_{\widehat\ell},
    \label{eq:theta_population_distribution}
\end{equation}
where $\mu$ is the population mean and $\tau^2$ is the between-individual biological variance. 
This normality assumption is primarily a convenient and interpretable first model for population heterogeneity. 
Other choices, such as log-normal, gamma, truncated normal, mixture, or fully nonparametric distributions, may be substituted when positivity, skewness, multimodality, or subpopulation structure is expected. 
In particular, for strictly positive quantities such as $r$ and $S_{\max}$, a log-normal random-effects model is a natural alternative.

To separate biological variability from estimation uncertainty, we approximate the uncertainty in each biological-parameter estimate. Let $\widehat{\Sigma}_{w}^{(i)}$ denote the covariance estimate for the WENDy coefficient vector $\widehat{\mathbf w}^{(i)}$. 
This covariance is computed from the weak-form generalized least-squares system as
\[
    \widehat{\Sigma}_{w}^{(i)}
    =
    \left[
        \left(\mathbf G^{(i)}\right)^\top
        \left(\mathbf R^{(i)}\right)^{-1}
        \mathbf G^{(i)}
    \right]^{-1},
\]
where $\mathbf G^{(i)}$ is the weak-form library matrix for trajectory $i$ and $\mathbf R^{(i)}$ is the corresponding weak residual covariance. 
The covariance is then propagated from coefficient space to biological-parameter space using the delta method \cite{VerHoef2012TheAmericanStatistician}. 
That is, if $    \mathbf J_i
    =
    D h_{\widehat\ell}
    \left(
        \widehat{\mathbf w}^{(i)}
    \right)$
is the Jacobian of the map from WENDy coefficients to biological parameters, then
\begin{equation}
    \widehat{\Sigma}_{\theta}^{(i)}
    =
    \mathbf J_i
    \widehat{\Sigma}_{w}^{(i)}
    \mathbf J_i^\top .
    \label{eq:delta_method_bio_covariance}
\end{equation}
For scalar shrinkage of a parameter $\theta$, we write $ s_i^2    = {\operatorname{Var}}
    \left(
        \widehat\theta^{(i)}
    \right).$
In other words, the raw WENDy biological-parameter estimate is modeled as
\begin{equation}
    \widehat\theta^{(i)}
    \mid \theta^{(i)}
    \sim
    \mathcal N(\theta^{(i)},s_i^2),
    \label{eq:observation_model_theta}
\end{equation}
while the true individual parameter satisfies the population model \eqref{eq:theta_population_distribution}, the estimate satisfies
\begin{equation}
    \widehat\theta^{(i)}
    \sim
    \mathcal N(\mu,\tau^2+s_i^2).
    \label{eq:marginal_theta_hat}
\end{equation}

The population parameters $\mu$ and $\tau^2$ are estimated by restricted maximum likelihood. 
That is, for a fixed value of $\tau^2$, we define the total variance as $V_i(\tau^2) := \tau^2+s_i^2$
and estimate the population mean as
\begin{equation}
    \widehat\mu(\tau^2)
    =
    \frac{
        \sum_{i\in\mathcal I_{\widehat\ell}}
        \widehat\theta^{(i)} / V_i(\tau^2)
    }{
        \sum_{i\in\mathcal I_{\widehat\ell}}
        1 / V_i(\tau^2)
    }.
    \label{eq:precision_weighted_mean}
\end{equation}
The restricted profile log-likelihood (up to an additive constant) is given by
\begin{equation}
    \ell_R(\tau^2)
    =
    -\frac12
    \left[
        \sum_{i\in\mathcal I_{\widehat\ell}}
        \log V_i(\tau^2)
        +
        \log
        \left(
            \sum_{i\in\mathcal I_{\widehat\ell}}
            \frac{1}{V_i(\tau^2)}
        \right)
        +
        \sum_{i\in\mathcal I_{\widehat\ell}}
        \frac{
            \left(
                \widehat\theta^{(i)}
                -
                \widehat\mu(\tau^2)
            \right)^2
        }{
            V_i(\tau^2)
        }
    \right].
    \label{eq:reml_profile_likelihood}
\end{equation}
We estimate $    \widehat\tau^2
    =
    \arg\max_{\tau^2\ge 0}
    \ell_R(\tau^2),$
and then set $    \widehat\mu    =  \widehat\mu(\widehat\tau^2).$
The estimated standard deviation $    \widehat\tau  =    \sqrt{\widehat\tau^2}$
is interpreted as the estimated biological between-individual variability in the parameter $\theta$, after accounting for WENDy estimation uncertainty.

The Bayes shrinkage posterior mean of the latent individual parameter is
\begin{equation}
    \widetilde\theta^{(i)}
    =
    \widehat\mu
    +
    \gamma_i
    \left(
        \widehat\theta^{(i)}-\widehat\mu
    \right),
    \qquad
    \gamma_i
    =
    \frac{\widehat\tau^2}
    {
        \widehat\tau^2+s_i^2
    }.
    \label{eq:eb_shrinkage}
\end{equation}
The factor $\gamma_i\in[0,1]$ is the shrinkage gain. 
When the WENDy estimate for individual $i$ is precise relative to the population variability, $s_i^2\ll \widehat\tau^2$, and $\gamma_i\approx 1$, so little shrinkage is applied. 
On the other hand, when the individual estimate is highly uncertain, $s_i^2\gg \widehat\tau^2$, and $\gamma_i\approx 0$, so the posterior mean is pulled strongly toward the population mean. 
The posterior variance of the individual parameter is
\begin{equation}
    \operatorname{Var}
    \left(
        \theta^{(i)}
        \vert\,
        \widehat\theta^{(i)}
    \right)
    =
    \frac{
        \widehat\tau^2 s_i^2
    }{
        \widehat\tau^2+s_i^2
    }
    =
    \gamma_i s_i^2 .
    \label{eq:posterior_variance_theta}
\end{equation}

It is important to distinguish the role of the raw WENDy estimates from the role of the empirical-Bayes shrinkage. 
The raw WENDy parameters $    \widehat{\mathbf w}^{(i)}$ are used as the individual-level fitted dynamics because they minimize the weak-form residual for each trajectory. 
The empirical-Bayes layer is instead used to estimate the population distribution of biological parameters and to correct the noise-inflated spread of the raw individual estimates. 
Thus, the primary population-level outputs are $    \widehat\mu,
    \,
    \widehat\tau^2,\,
    \left\{s_i^2\right\}_{i\in\mathcal I_{\widehat\ell}},
    \, \text{and }
    \left\{\gamma_i\right\}_{i\in\mathcal I_{\widehat\ell}}.$
The shrunken quantities $\widetilde\theta^{(i)}$ are interpreted as empirical-Bayes posterior summaries of the parameter distribution, not necessarily as improved individual trajectory-level fits.
This distinction is important because shrinkage may worsen the fit to particular individuals, especially individuals whose true growth parameters are far from the population mean, while still improving the estimation of the population-level distribution.

\section{Results}\label{sec:results}

\subsection{Synthetic data}\label{ssec:synthetic}
We evaluate the proposed method on synthetic ensembles generated from each of the five candidate growth laws in Table~\ref{tab:growthModels}, with controlled heterogeneity and additive measurement noise. 
The synthetic study isolates the three quantities the method aims to recover from data: the functional form of $g$, the individual-level coefficients $\{\mathbf{w}^{(i)}\}$, and the population-level heterogeneity $\tau^2$ in the biological parameters $r$ and $S_{\max}$.

\subsubsection{Test models}

To assess the generality of the method, we consider five canonical growth laws that span a range of qualitative behaviors which are presented in \Cref{tab:growthModels}.
While this set of models spans many biological applications, it is by no means complete and the members of this set of models should be adjusted based on the application at hand.

\begin{table}[h]
\centering
\small
\begin{tabular}{@{}llll@{}}
\toprule
Model Name & Usual r.h.s. & WENDy r.h.s. & Example systems \\
\midrule
Exponential & $rx$ & $w_1 x$ & Cell growth \cite{RobertHoffmannKrellEtAl2014BMCBiol,CadartVenkovaPielEtAl2022eLife} \\
Logistic & $rx(1 - x/S_{\max})$ & $w_1 x + w_2 x^2$ & Plant height \cite{SunFrelich2011JEcol} \\
Gompertz & $rx \log(S_{\max}/x)$ & $w_1 x + w_2 x \log x$ & Tumor growth \cite{Laird1964BrJCancer}; Oocyte dynamics \cite{FostierLesageThermesEtAl} \\
Linear von Bertalanffy & $r(S_{\max} - x)$ & $w_1 + w_2 x$ & Fish length \cite{BevertonHolt1993,VincenziJesensekCrivelli} \\
Metabolic von Bertalanffy & $r(S_{\max}^{1/3} x^{2/3} - x)$ & $w_1 x + w_2 x^{2/3}$ & Allometric body-size growth \\
\bottomrule
\end{tabular}
\caption{\label{tab:growthModels}Candidate growth laws used for WENDy-based model selection.
$x$ denotes individual size, $r > 0$ is the intrinsic growth rate, and
$S_{\max} > 0$ the asymptotic size. The third column gives the
linear-in-coefficients form of the right-hand side (r.h.s.) used in weak-form regression.}
\end{table}

For each model in \Cref{tab:growthModels}, we generate $I=500$ synthetic trajectories as follows:  
The biological parameters
$\theta^{(i)} = (r^{(i)}, S_{\max}^{(i)})$ are drawn i.i.d. from a multivariate Gaussian whose mean is set to a biologically reasonable value for the model in question and whose covariance is chosen to produce a coefficient of variation of $10\%$ in each component.
Each trajectory is obtained by integrating the corresponding ODE with a fourth-order Runge--Kutta scheme on a uniform grid until a ``death event'' occurs, after which independent Gaussian measurement
noise is added with signal-to-noise ratio $\sigma_{\mathrm{nr}}$.
For more details on how the synthetic data is generated see Appendix~\ref{asec:ArtificialSims}.

There are three properties of this protocol worth pointing out. 
First, the heterogeneity is built in to the parameters, not to the dynamics. 
In other words, each individual obeys the deterministic ODE \eqref{eq:growth_ode} exactly and all observed spread in the trajectories is either a consequence of parameter spread or of measurement noise.
Second, all initial sizes are drawn from a narrow Gaussian distribution, so the method must succeed without a common reference point between the trajectories.
Finally, we do not perturb $\theta^{(i)}$ over the course of a trajectory.
This of course is possible using time-varying parameter estimation methods, but as this has yet to be developed for WENDy, we leave this to future work.

\subsubsection{Model Selection}
\begin{figure}[h]
    \centering
    \includegraphics[width=0.75\linewidth]{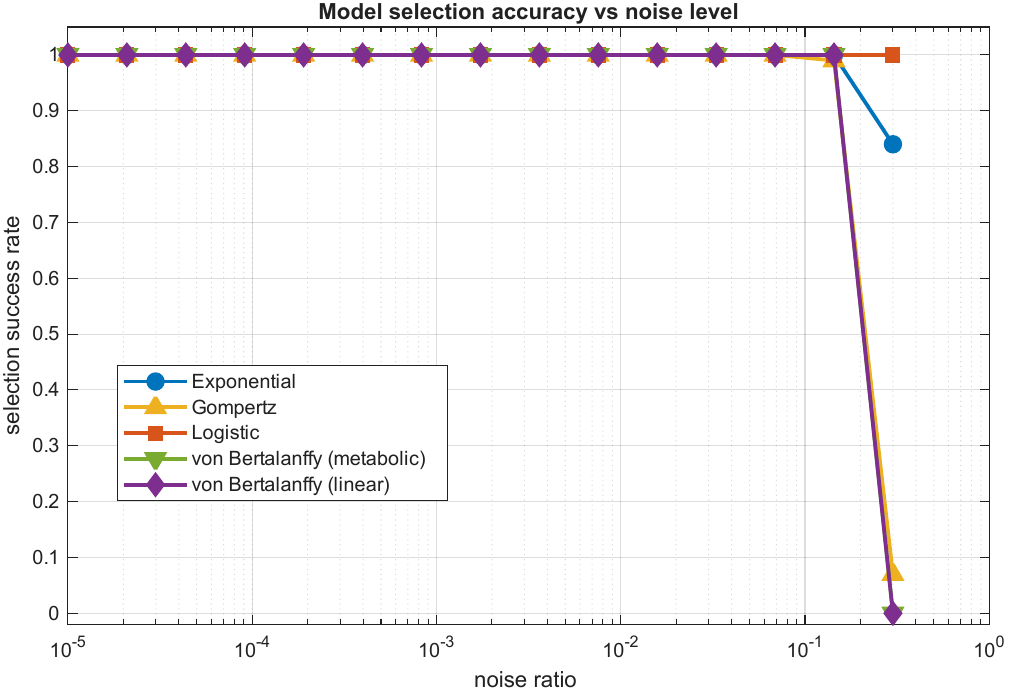}
    \caption{Mean success rate of the model selection process described in \cref{subsec:wendy_model_comparison} as a function of the additive measurement noise. The mean is taken over 100 realizations of each noise level. }
    \label{fig:sucessrate}
\end{figure}

We quantify the performance of the BIC model selection step via the selection success rate (i.e., the rate at which the true model was successfully identified) of model selection over 100 realizations of different noise levels.
In Figure~\ref{fig:sucessrate}, we plot the success rate as a function of the noise level.
We see that up until about $\approx 25\%$ noise, the method is correctly able to select the true model. 
At this threshold, distinguishing features of the model such as location of inflection points and existence of asymptotic sizes are no longer visible in the data, resulting in similar BIC scores between all models.

\subsubsection{Parameter recovery}
\begin{figure}[h]
    \centering
    \includegraphics[width=0.5\linewidth]{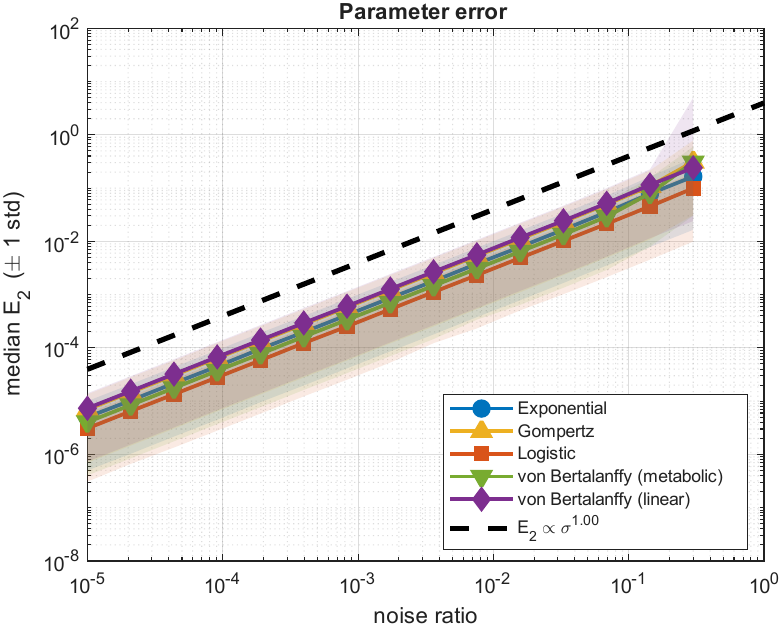}
    \caption{Median $E_2$ error and standard deviation for the five artificial data sets from the models in Table~\ref{tab:growthModels} over various noise levels. This median is taken over 100 realizations of each noise level.}
    \label{fig:E2_vs_noise}
\end{figure}

We quantify per-trajectory parameter accuracy via the relative $\ell_2$-error
\begin{equation}  \label{eq:E2}
  E_2\!\left(\widehat{\mathbf{w}}^{(i)}\right)
  = \frac{\|\widehat{\mathbf{w}}^{(i)} - \mathbf{w}^{(i)}_{\text{true}}\|_2}
         {\|\mathbf{w}^{(i)}_{\text{true}}\|_2},
\end{equation}
and report the median over individuals together with the
$\pm 1$ standard deviation envelope across Monte Carlo realizations in Figure~\ref{fig:E2_vs_noise}.
This is a standard performance metric in parameter identification and in particular the WENDy framework \cite{BortzMessengerDukic2023BullMathBiol,RummelMessengerBeckerEtAl2025arXiv250208881,Heitzman-BreenDukicBortz2026BullMathBiola}.
Across all five models the median $E_2$ grows linearly with the noise ratio, $E_2 \propto \sigma_{\mathrm{nr}}$.
This linear scaling is the expected behavior of an asymptotically consistent weak-form estimator under sub-noise-level perturbations, and matches the theoretical predictions of \cite{MessengerBortz2024IMAJNumerAnal,BortzMessengerDukic2023BullMathBiol}.

\begin{figure}[h!]
    \centering    \includegraphics[width=0.95\linewidth,
    height = 0.4\linewidth]{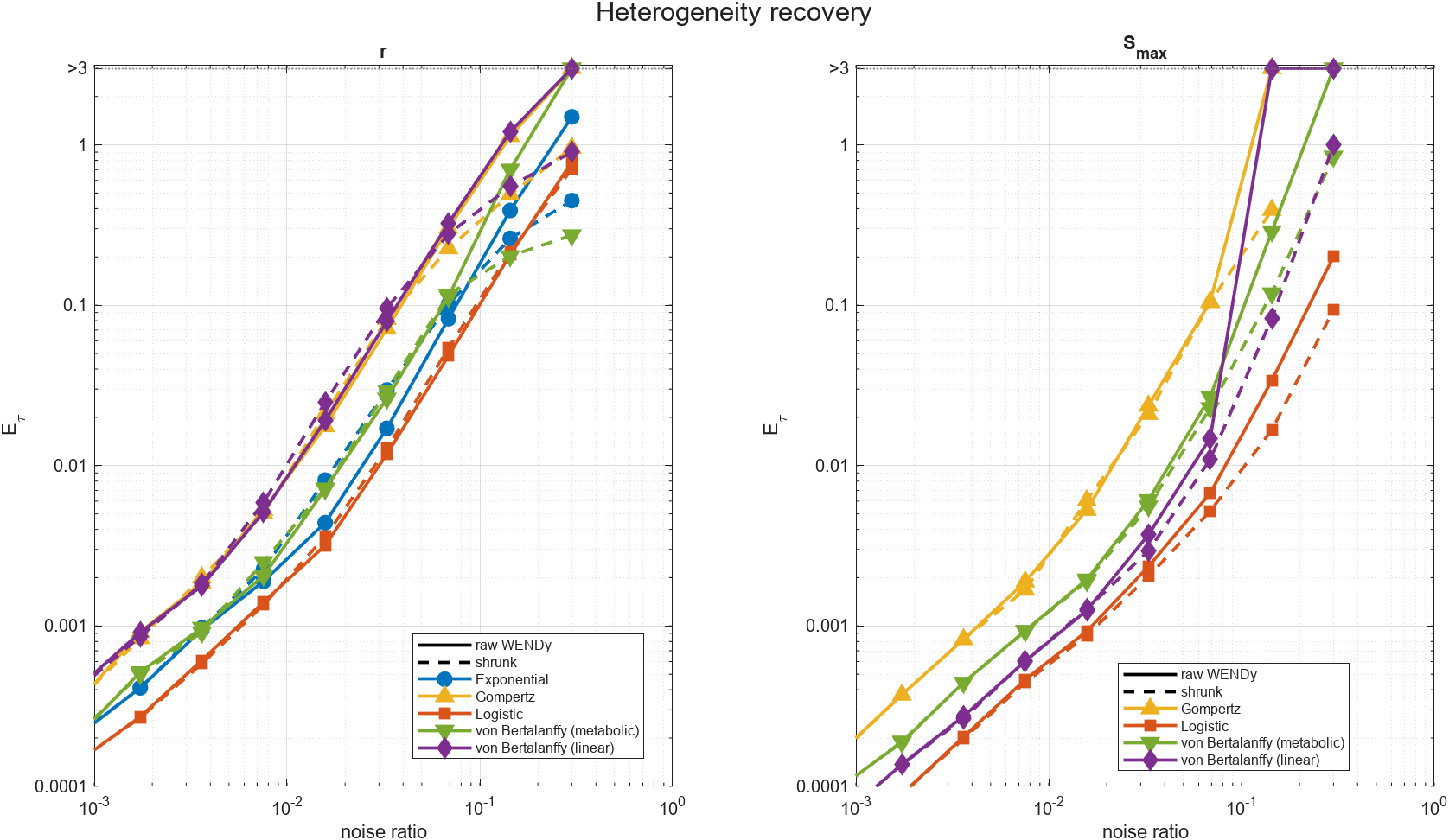}
    \caption{$E_\tau$ trend as measurement noise increases for the biological parameters $r$ and $S_\text{max}$. 
    Note that at higher noise levels, the raw WENDy estimates can substantially over estimate the true variance. On the other hand, the shrinkage method collapses all of the estimates to the mean parameter resulting in an error of 1.}
    \label{fig:E_tau_vs_noise}
\end{figure}

We also report the relative error in the recovered between-individual
standard deviation,
\begin{equation}\label{eq:E_tau}
  E_\tau(\widehat{\tau})
  = \frac{|\widehat{\tau} - \tau_{\text{true}}|}{|\tau_{\text{true}}|},
\end{equation}
for each biological parameter ($r$ for all laws, and $S_{\max}$ for the four saturating laws), and compare the estimated heterogeneity from the raw WENDy estimates and the shrunken estimates in Figure~\ref{fig:E_tau_vs_noise}.
We show only the results for noise levels above $\sigma_{\mathrm{nr}}>10^{-3}$ since, as expected, for exceptionally low noise, the estimators are close to identical implying the raw WENDy estimates can accurately recover the heterogeneity without any added post processing. 
However, at higher noise levels, the raw estimates increasingly overestimate heterogeneity, requiring the empirical Bayes shrinkage correction.
We do point out that, depending on the estimated noise, the shrinkage correction can over correct causing an under representation of the heterogeneity. 
Finally, at exceptionally high noise levels (e.g., $\sigma_{\mathrm{nr}}\geq 0.15$), both estimators degrade, but in different ways. 
The raw WENDy estimate strongly overshoots the true heterogeneity.
On the other hand, the shrinkage estimate saturates the $E_\tau$ error at 1, representing the shrinkage
gains $\gamma_i$ have collapsed to zero and every individual estimate is pulled to the population mean.
In this regime, the data is simply insufficient to resolve heterogeneity from noise.

\subsection{Real Data}
In this subsection, we apply the method to openly available experimental data sets and explore how the results of the method compare to the results of previous studies.  

\subsubsection{\textit{E. coli} Cells}\label{subsec:Ecoli_example}

We applied the above method to single-cell length trajectories of \emph{E.\ coli} cells grown in three M9 minimal media (alanine, glycerol, and glucose-casamino-acid) at $28\,^\circ$C, from the openly available dataset of \cite{KarTiruvadi-KrishnanMannik2021eLife} hosted on \cite{DVN/BNQUDW_2021}.
Length trajectories were obtained by phase-contrast microscopy in a mother-machine device at $4$-minute frame rate, with each trajectory covering one cell cycle from birth to division. 
The mean reported generation times in the three media are $\langle T_d \rangle = 214$, $164$, and $65$ minutes, and the corresponding dataset sizes are $N_{\text{tot}} = 816$, $648$, and $737$ cell cycles. 
We subsample uniformly to $N = 500$ trajectories per medium for direct cross-medium comparability.
We also apply minor pruning of trajectories (those with less than 6 observations) and truncate any trajectory with sharp decline (over 0.4 $\mu$m in one time frame) as these appear to be mostly tracking errors on a small subset of trajectories.

\begin{figure}[h]
    \centering
    \includegraphics[width=\linewidth]{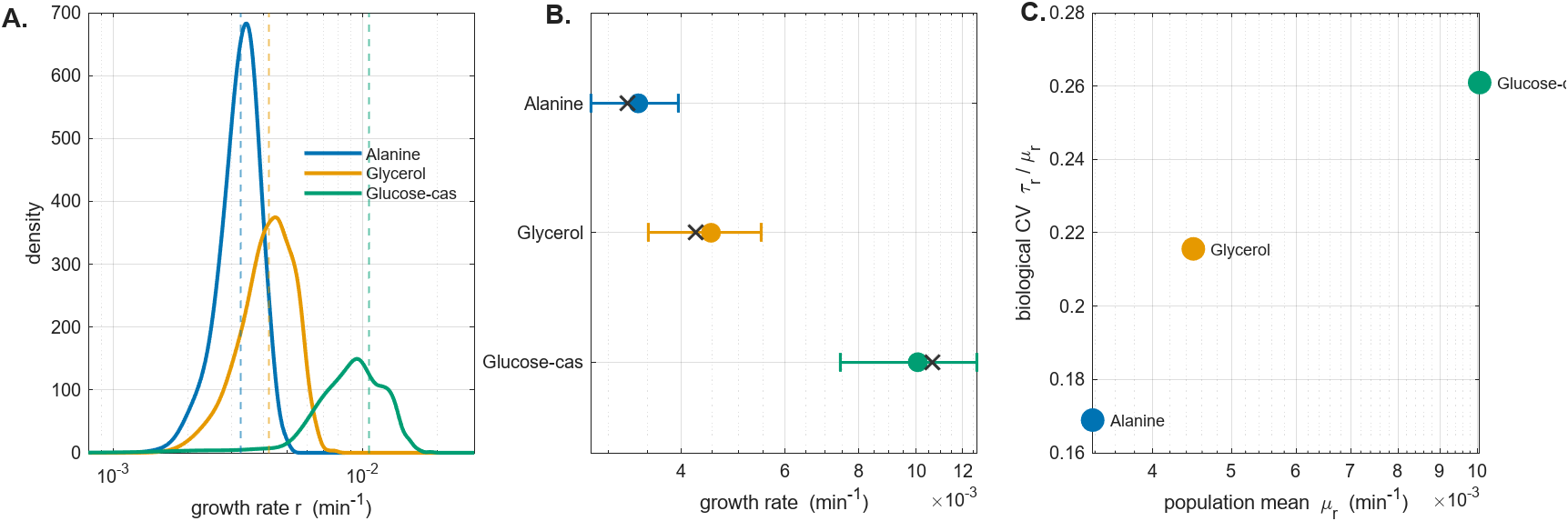}
    \caption{Comparison of the recovered single-cell \emph{E.~coli} growth rates for the different media (alanine (blue), glycerol (orange), and glucose-casamino acid (green)) using the size trajectory data from \cite{KarTiruvadi-KrishnanMannik2021eLife}.
\textbf{A:} Population distributions of the intrinsic growth rate $r$; dashed vertical lines mark the estimate of each medium's mean growth rate reported in \cite{KarTiruvadi-KrishnanMannik2021eLife}.
\textbf{B:} Population mean $\hat{\mu}_r$ (circles) and biological heterogeneity $\pm \hat{\tau}_r$ (error bars), compared with the values reported in \cite{KarTiruvadi-KrishnanMannik2021eLife} (black~$\times$). 
Error bars represent population heterogeneity in $r$, not standard error of the mean.
\textbf{C:} Biological coefficient of variation $\hat{\tau}_r / \hat{\mu}_r$ as a function of the population mean growth rate. 
All panels use $N = 500$ trajectories per medium, subsampled uniformly from the full dataset.}
    \label{fig:ecoli_media}
\end{figure}

The summed per-trajectory BIC criterion of Section~\ref{subsec:wendy_model_comparison} strongly prefers the exponential law for the three data sets with
 population means $\widehat{\mu}_r = 3.38\!\times\!10^{-3}$,
$4.49\!\times\!10^{-3}$, and $1.01\!\times\!10^{-2}\,\mathrm{min}^{-1}$ for alanine, glycerol, and glucose-cas, respectively (Figure~\ref{fig:ecoli_media}). 
Each agrees with the values of  $\ln 2 / \langle T_d \rangle$ reported in \cite{KarTiruvadi-KrishnanMannik2021eLife} to approximately $4$--$6\%$ (alanine $+4.3\%$, glycerol $+6.2\%$, glucose-cas $-5.6\%$).
The recovered biological coefficients of variation are $\widehat{\tau}_r / \widehat{\mu}_r = 0.17$, $0.22$, and $0.26$, respectively.

\subsubsection{HeLa Cells}
The HeLa cell volume growth dataset of \cite{CadartVenkovaPielEtAl2022eLife} (openly available on Dryad \cite{CadartPielCosentinoLagomarsino2022})  comprises 1,696 single-cell volume trajectories measured by fluorescence exclusion microscopy (FXm) at 10-minute intervals across four independent replicates. 
HeLa cells expressing the hgeminin-GFP cell cycle reporter were imaged asynchronously and each trajectory was partitioned into G1 and S-G2 segments. 
To characterize the growth mode, the original authors computed discrete volume derivatives $\frac{\diff{V}}{\diff{t}}$ over 50-minute sliding windows, binned the resulting growth speeds by instantaneous volume, and fit linear slopes to the population-averaged $\frac{\diff{V}}{\diff{t}}$ versus $V$ relationship, yielding estimates of the volume-specific growth rate $r \approx 0.038–0.047\, \text{hr}^{-1}$. 
When this analysis was stratified by cell cycle phase, the authors found that the specific growth rate in S-G2 exceeds that in G1 by approximately 15\%, with each phase exhibiting a rate that is approximately constant across cell volumes.

\begin{figure}[ht]
    \centering    \includegraphics[width=0.45\linewidth,height = 0.4\linewidth]{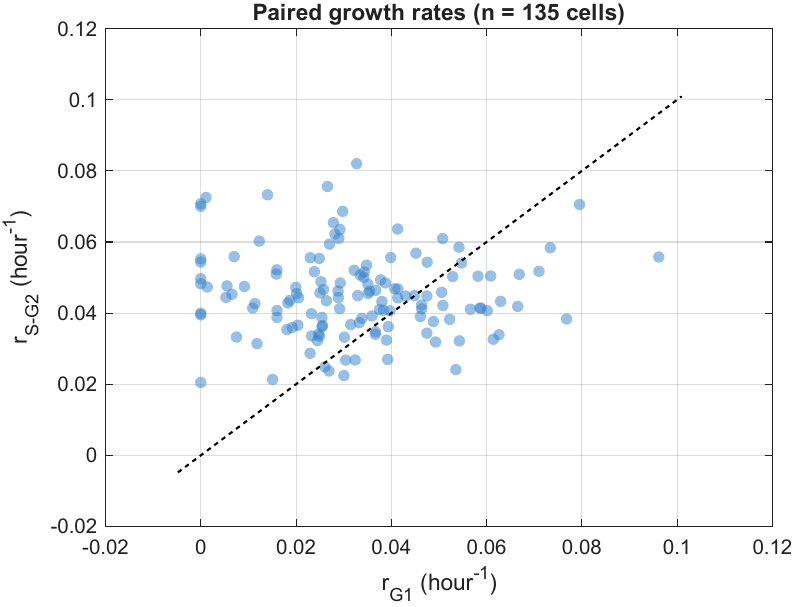}    \includegraphics[width=0.45\linewidth,height = 0.4\linewidth]{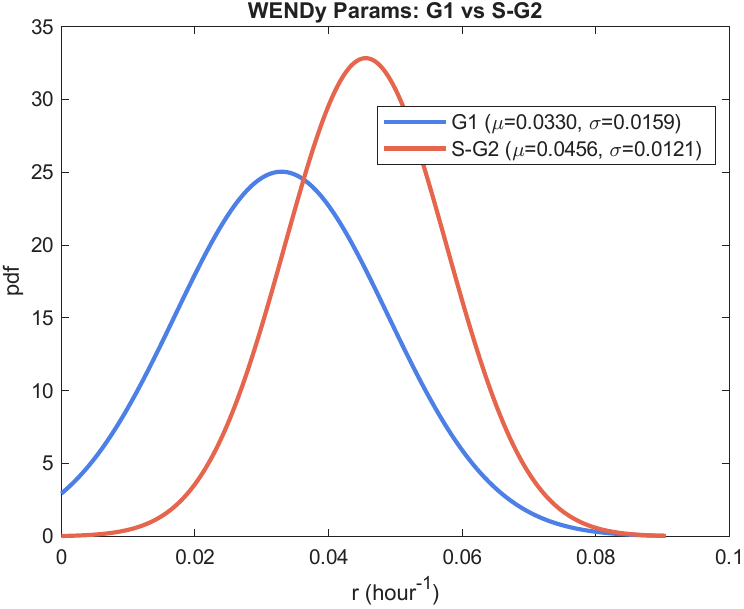}
    \caption{\textbf{Left:} Scatter plot of the estimated growth rates of HeLa cells observed in both phases of growth against the line $y=x$.
    \textbf{Right:} Normalized estimated parameter population distributions of the two growth phases.}
    \label{fig:HeLaParams}
\end{figure}

Applying our WENDy-based method to their data confirms these findings. 
The BIC model selection unambiguously identifies exponential growth ($\frac{\diff{V}}{\diff{t}} =rV$) as the best-supported law in both phases, outperforming the two-parameter logistic, Gompertz, and von Bertalanffy models. 
The WENDy estimated population-mean growth rates are $r_{\text{G1}}\approx 0.033 \,\text{hr}^{-1}$ and $r_{\text{SG2}}\approx 0.046 \,\text{hr}^{-1}$. 
The S-G2 rate falls within the aforementioned work's estimated 0.038–0.047 range, while the G1 rate sits slightly below it.
This discrepancy is most likely due to the G1 trajectories including the first  $\approx$1.3 hours after cytokinesis, during which the authors observed a non-exponential phase, which they removed from their growth-rate estimates.
The WENDy results are presented in Figure~\ref{fig:HeLaParams} where we present a scatter plot of learned parameter values for those 135 cells observed in both phases of growth and the parameter distribution over the population.
Matching the observations reported in \cite{CadartVenkovaPielEtAl2022eLife}, we see a significant increase in the mean growth rate estimate between the G1 and S-G2 phases.


\subsubsection{Loblolly Pines}
\begin{figure}
    \centering
    \includegraphics[width=0.85\linewidth]{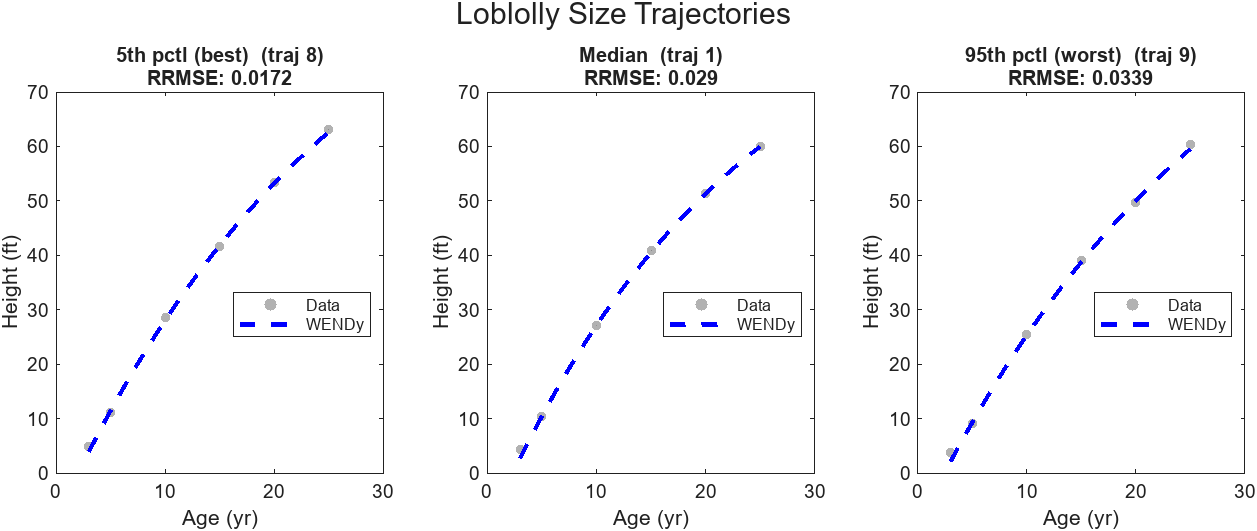}
    \caption{Example linear von Bertalanffy trajectories (blue dashed) selected and fit with the WENDy method for the sparse Loblolly Pine data (gray dots).}
    \label{fig:LoblollyTraj}
\end{figure}

In this example, we make use of a Loblolly pine tree size dataset \cite{kung1986fitting,pinheiro2000mixed}.
This dataset consists of height measurements of 14 \textit{Pinus taeda} trees, each grown from a distinct seed source and whose heights were measured at ages 3, 5, 10, 15, 20, and 25 years. 
Heights of the trees range from approximately 4 to 60 feet, spanning the transition from early exponential-like growth into the decelerating phase where trunk diameter and height approach asymptotic limits. 
This data set is generally used as a standard benchmark for nonlinear mixed-effect models since its inclusion in R's \texttt{datasets} package.
For our purposes, this data set presents an interesting challenge for WENDy as the data is relatively sparser than datasets typically used in weak-form benchmarks, causing a problem when calculating the numerical quadrature. 
This can be solved via linear interpolation of the data points to ensure adequate resolution for the weak-form test function integrals in the WENDy method (see \cite{heitzman2026weak} for an exploration of WENDy with interpolated data).
We point out that interpolation does not add information of the dynamics to WENDy, but simply allows us to fill out the numerical quadrature.
With these interpolated trajectories, the BIC model selection identifies the linear von Bertalanffy model ($\frac{\diff{h}}{\diff{t}} = r(S_{\max} - h)$), consistent with the asymptotic
regression model used by Pinheiro and Bates \cite{pinheiro2000mixed} in their treatment of this dataset, and in contrast to the logistic model originally fit by Kung \cite{kung1986fitting}.
The 5th percentile, median, and 95th percentile trajectory fits as well as their root relative mean squared error (RRMSE) are provided in Figure~\ref{fig:LoblollyTraj}.

\section{A discussion on upscaling to the population-level}\label{sec:population_level}
 While the method above provides a robust model selection and parameter distribution estimation method, it remains to show how one progresses from these estimates, $(g, \widehat{\mathcal{P}})$, to population level predictions. 
 To this end, let $u(t,x,\theta)$ denote the density of individuals with size $x$ and parameter vector $\theta$ at time $t$. 
 Ignoring source and sink processes such as birth, death, and cell division\footnote{In the case that these dynamics are homogeneous across the population, their inclusion falls naturally. The more realistic and more complicated case of additional heterogeneous parameters affecting these dynamics is left as future work.}, the deterministic dynamics \eqref{eq:growth_ode} imply that $u$ follows the continuity equation
 \begin{equation}
    \partial_t u(t,x,\theta)
    +
    \partial_x\!\bigl(g(x;\theta)\,u(t,x,\theta)\bigr)
    =
    0.
    \label{eq:joint_transport}
\end{equation}
This equation describes conservation of individuals in the joint state-parameter space.
The observable population density in size alone is the marginal
\begin{equation}
    \rho(t,x)
    =
    \int u(t,x,\theta)\diff{\theta},
    \label{eq:marginal_density}
\end{equation}
and integrating \eqref{eq:joint_transport} with respect to $\theta$ yields
\begin{equation}  \label{eq:exact-effective}
  \partial_t \rho(t, x) + \partial_x \bigl(g_{\mathrm{eff}}(t, x)\,\rho(t, x)\bigr) = 0,
  \qquad
  g_{\mathrm{eff}}(t, x) := \mathbb{E}\bigl[g(x; \theta)\mid x, t\bigr]
  = \frac{\int g(x; \theta)\, u(t, x, \theta)\,\mathrm{d}\theta}{\rho(t, x)}.
\end{equation}
Equation~\eqref{eq:exact-effective} is exact, but not closed in $\rho$ alone: $g_{\mathrm{eff}}$ depends on the conditional law of $\theta$ given size $x$ at time $t$, which is generally not equal to the marginal $\mathcal{P}(\theta)$ because the dynamics induce a correlation between $\theta$ and $x$. 
The remainder of this section discusses two complementary ways to approximately close \eqref{eq:exact-effective} from the outputs of WENDy.

\subsection{The mixture-of-cohorts closure}
Following the work of \cite{BanksFitzpatrick1991QuartApplMath}, we can discretize the parameter distribution by $M$ representatives $\{\theta^{(m)}\}_{m=1}^M$ with corresponding weights $\{\alpha^{(m)}\}_{m=1}^M$ satisfying $\sum_m \alpha_m = 1$.
Then one can approximate the density as a sum of cohorts:
\begin{equation}
  u(t, x, \theta) \;\approx\; \sum_{m=1}^M \alpha_m\, u^m(t, x)\,\delta\!\bigl(\theta - \theta^{(m)}\bigr),
\end{equation}
where each cohort density $u^m$ satisfies a one-dimensional transport equation with its own deterministic growth rate
\begin{equation} \label{eq:cohort-pde}
  \partial_t u^m(t, x) + \partial_x\!\bigl(g(x; \theta^{(m)})\, u^m(t, x)\bigr) = 0,
  \qquad m = 1, \dots, M.
\end{equation}
The population density is approximated as
$\rho(t, x) \approx \sum_m \alpha_m\, u^m(t, x)$.

\begin{figure}[tbp]
    \centering
    \includegraphics[width=0.48\linewidth]{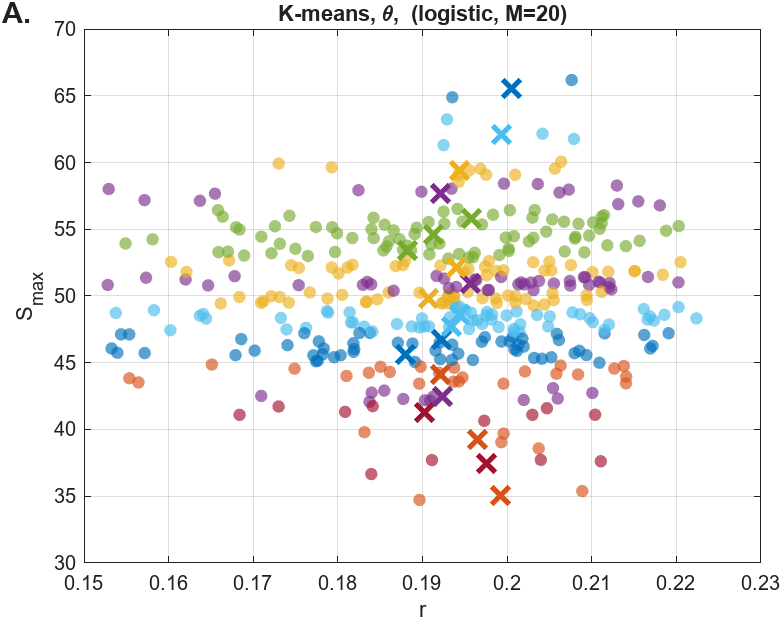}
    \includegraphics[width=0.48\linewidth]{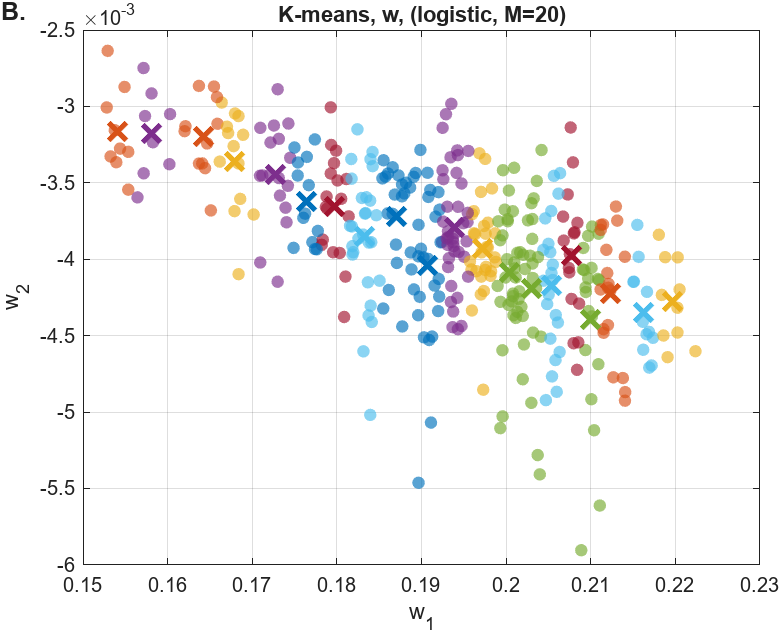}
    \includegraphics[width=0.96\linewidth]{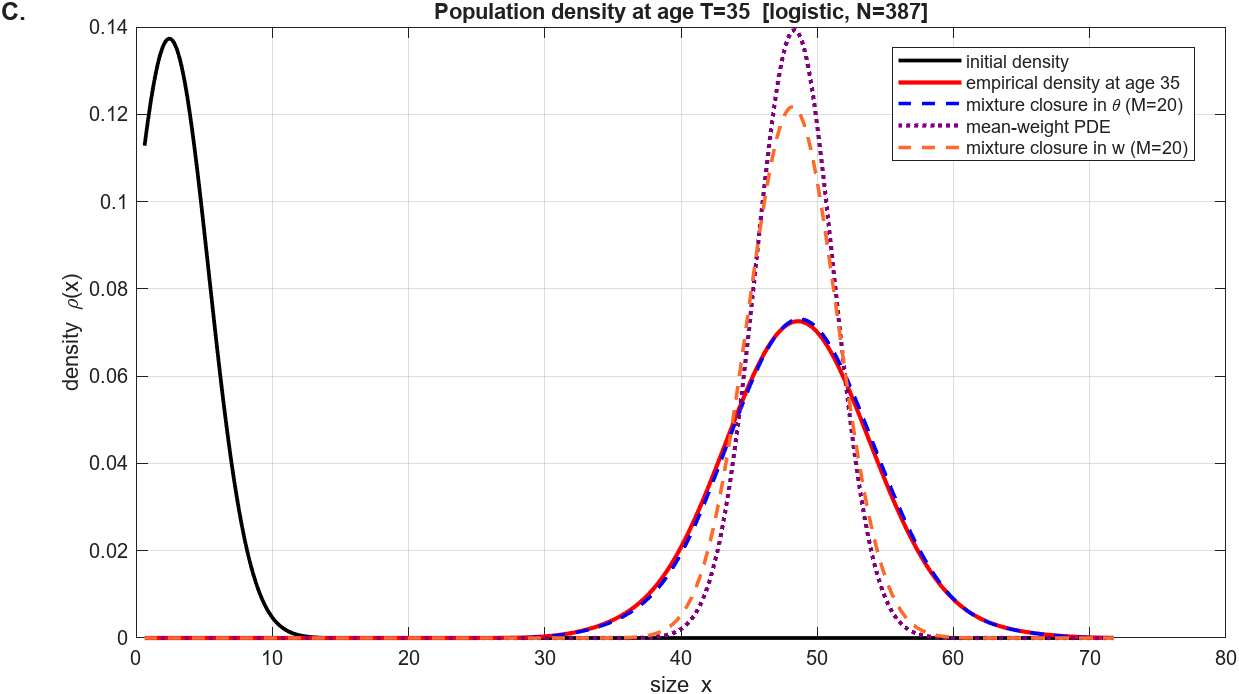}
    \caption{\textbf{A \& B:} K-means clustering algorithm applied to the learned parameter space for the artificial logistic data set with $\sigma_{\mathrm{nr}} = 0.05$. Plot A shows the k-means algorithm applied to the biological parameter space $(r,S_{\max})$ and plot B to the LiP space $(w_1,w_2)$. Coloring represents the cohort groups with crosses ($\times$) representing the representative of the nearby cohort.
    \textbf{C:} Population-at-age density simulations using the clustering results above. 
    Solid lines represent empirical densities from the data, with the black line representing the initial density and the red line representing the size density of individuals at age 35. The dotted ($\cdots$) line represents the simulation using the mean parameter whereas the dashed (-\,-) lines use the results of the two clustering algorithms.}
    \label{fig:kmeans_logisitc}
\end{figure}

\begin{figure}
    \centering
    \includegraphics[width=0.75\linewidth]{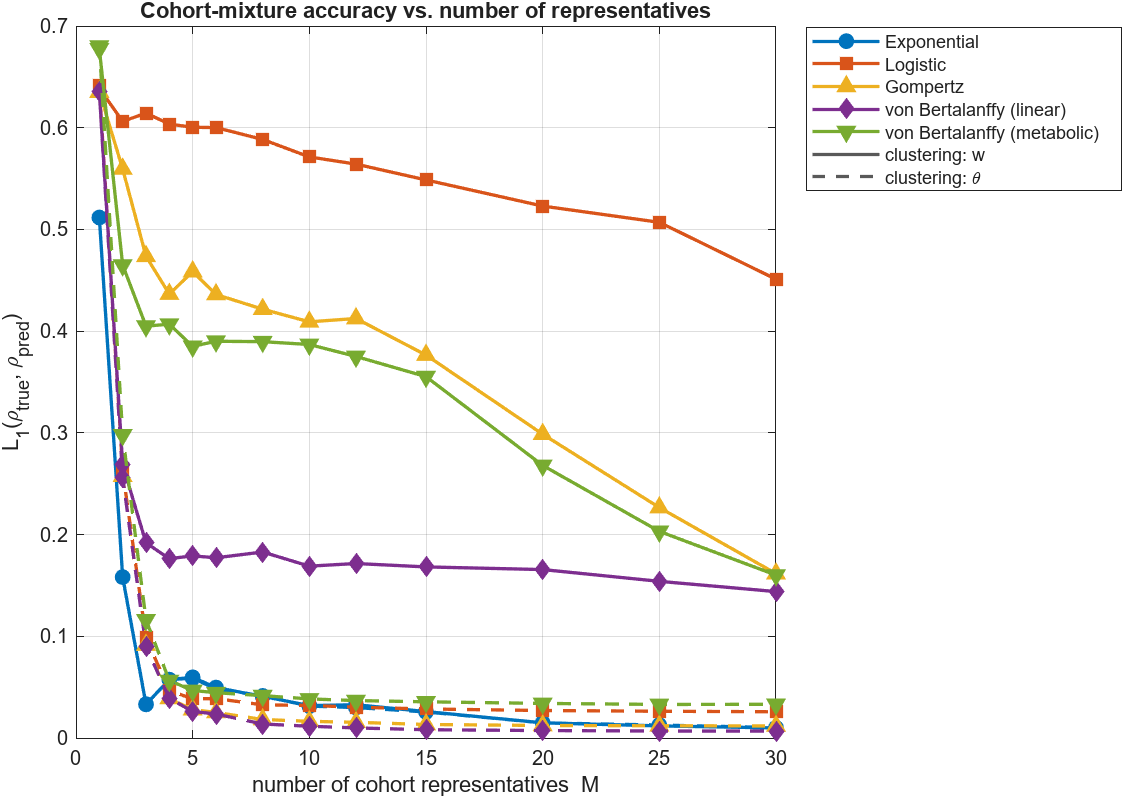}
    \caption{$L_1$ accuracy of the cohort mixture method as a function of the number of cohort representatives $M$, for each growth law fit to clean data. Solid lines represent the clustering done in LiP space and dashed lines represent the biological parameter space.}
    \label{fig:L1_vs_Cohorts}
\end{figure}

For the method of choosing representatives for simulation/prediction, any quadrature rule on the estimated parameter distribution $\widehat{\mathcal{P}}$ yields valid representative parameters and weights.
For a natural data-driven choice, we recommend clustering the individual parameter estimates and taking $\theta^{(m)}$ (or alternatively $\mathbf{w}^{(m)}$) as the centroid of cluster $m$ and $\alpha_m$ as its mass fraction.
In Figure~\ref{fig:kmeans_logisitc}\textbf{A\&B}, we use the k-means clustering algorithm for both the biological parameter space and LiP parameter space for $M=20$ groups. 
We generally find that fewer cohort groups are needed (for the same accuracy) when the clustering is done on the biological parameters $\theta = (r,S_{\max})^T$ rather than in the weight space given by $\mathbf{w}$.
This is demonstrated in Figure~\ref{fig:kmeans_logisitc}\textbf{C} where we simulate size-at-age population densities with k-means clustering applied in both parameter spaces and is confirmed on all models in Figure~\ref{fig:L1_vs_Cohorts} where we compare the $L_1$ norm, $L_1(\rho_{\text{true}}(T,\cdot),\rho_{\text{pred}}(T,\cdot)) = \int|\rho_{\text{true}}(T,x) -\rho_{\text{pred}}(T,x)|\diff{x}$, for the true density and predicted density as a function of the number of cohort representatives.
Notice that performing the clustering algorithm in the biological parameter space greatly out performs clustering in the LiP space.

\subsection{Effective-velocity closure}
The simplest closure to equation~\eqref{eq:exact-effective} is to simply replace the conditional expectation with an unconditional approximation:
\[g_{\text{eff}}(x) \approx \mathbb{E}_{\theta \sim\widehat{\mathcal{P}}}[g(x;\theta)],\]
and make use of the single resulting transport equation.
For laws linear in $\mathbf{w}$ this reduces to using $g(x;\bar{\mathbf{w}})$ where $\bar{\mathbf{w}}$ is the mean of the WENDy weights (see Figure~\ref{fig:kmeans_logisitc}\textbf{C}).
This closure is computationally more efficient than the cohort-mixture approach as, for predictions or simulations, one only needs to solve a single PDE as oppose to the $M$ weighted PDEs \eqref{eq:cohort-pde}. 
However, this approximation cannot reproduce the correct long-time asymptotics and is, therefore, a short-time approximation. 
For example, for every saturating candidate model in Table~\ref{tab:growthModels}, each individual approaches its own asymptotic size, i.e., $x^{(i)}(t) \to S_{\max}^{(i)}$ as $t\to \infty$. 
The long-time marginal density (again ignoring source and sink processes) is therefore the pushforward of the parameter distribution through the transport map:
\[\rho_\infty(x) = \int \delta(x-S_{\max}(\theta)) \diff{\mathcal{P}}(\theta),\]
which is (generally) a non-degenerate distribution.
On the other hand, the effective velocity closure uses a single effective maximum size $\bar{S}_{\max}$ in the dynamics and therefore the long-time limit of the population density is to concentrate at this effective maximum size, e.g., $\rho_\infty(x) = \delta(x- \bar{S}_{max})$.

\section{Discussion}\label{sec:discussion}

In conclusion, we have presented a WSciML method for growth law selection and parameter identification of heterogeneous parameters in a population of growing individuals.
The  method combines the WENDy method for parameter identification, BIC model selection, and empirical Bayes shrinkage to correct for artificial variance due to measurement noise. 
We tested the method on several artificial data sets to explore its performance at different noise regimes.
From Figure~\ref{fig:sucessrate}, it is clear that the model selection struggles at high noise ($\sigma_{\mathrm{nr}}\approx 25\%$) at which point the identifying features of the models become indistinguishable from noise.
In such a high noise regime, it is better to fix an interpretable model from either first principles or from literature reference. 
It is worth noting that the parameter estimation step can still succeed in this regime for a fixed model. 
Indeed from Figure~\ref{fig:E2_vs_noise}, we see the trend of the median parameter error continues to grow proportional to the noise level. 
This is a commonly encountered strength of weak-form approaches due to all derivatives being removed from the noisy state variable and has been reported in many studies \cite{BortzMessengerDukic2023BullMathBiol,Heitzman-BreenDukicBortz2026BullMathBiola,MessengerBortz2021JComputPhys,MessengerBortz2021MultiscaleModelSimul,RummelMessengerBeckerEtAl2025arXiv250208881,TranBortz2025arXiv250703206b}.

The results of this work are a promising first step to the full inference of individual to population level heterogeneous growth dynamics. 
There are still multiple open questions to the work discussed here which are active frontiers.
First, this study is primarily focused on growth rates and leaves the treatment of death, division, birth, and other processes as future work; however, these processes are critical for prediction at the population level.
Additionally, the rigorous convergence of the cohort method discussed in Section~\ref{sec:population_level} to the population density described by \eqref{eq:joint_transport} (including source terms) is, to our knowledge, an open question.
Results concerning similar methods in the traditional setting of structured population models (i.e., when each individual follows the same characteristic equations) such as the Escalator Boxcar Train (EBT) and splitting-particle methods have been well established \cite{Carrillo.etal2019SIAMJNumerAnal,Gwiazda.etal2023IMAJNumerAnal,CarrilloGwiazdaUlikowska2014MathModelsMethodsApplSci} and the techniques used for these methods should be applicable to this problem.
Third, regarding the \textit{E. coli} example of Section~\ref{subsec:Ecoli_example}, the original work \cite{KarTiruvadi-KrishnanMannik2021eLife} concludes that the recorded trajectories could be super-exponential, however, none of the models implemented in this work describe this behavior and the more traditional growth model for bacteria cells was selected. 
It is possible that, if given the option, the model selection stage of the method (or an equation learning algorithm like WSINDy) could select a super-exponential model over the exponential model in our work.  
Finally, the development of a full mixed-effects-aware WENDy method, where the population mean and variance are used directly in the WENDy parameter estimation, could substantially improve the method presented here.

\paragraph{Data Availability:}
All code used in this manuscript will be made available upon publication on the repository \href{https://github.com/MathBioCU/Heterogeneous_Growth_Laws}{github.com/MathBioCU/Heterogeneous\_Growth\_Laws}.\\ 

\paragraph{Acknowledgments:}
{The research reported in this publication was supported in part by the NIGMS Division of Biophysics, Biomedical Technology and Computational Biosciences (grant R35GM149335 to DMB); NSF Division of Molecular and Cellular Biosciences MODULUS (grant 2054085 to DMB); NSF Division Of Environmental Biology (EEID grant DEB-2109774 to VD), and NIFA Biological Sciences (grant 2019-67014-29919 to VD). 
The funders had no role in study design, data collection and analysis, decision to publish, or preparation of the manuscript.}
This work utilized the Blanca condo computing resource at the University of Colorado Boulder.
Blanca is jointly funded by computing users and the University of Colorado Boulder.
The authors would like to thank  N.~Heitzman-Breen (CU Boulder) for useful discussion regarding mixed-effect modeling and would also like to thank M. Doumic (Sorbonne Universit\'{e}s, Inria, \'{E}cole polytechnique, CMAP) for insightful discussion about heterogeneity in growth rates for cell populations.

\bibliographystyle{siam}
\bibliography{bib}

\appendix
\section{Artificial Simulations}\label{asec:ArtificialSims}
For each growth law in Table~\ref{tab:growthModels} we generate $I = 500$ independent trajectories by the following procedure.
For each individual $i= 1,2,\dots,I$, the biological parameters $r^{(i)}$ and $S^{(i)}_{\max}$ ($r^{(i)}$ only in the case of exponential growth) are drawn independently from positive Gaussian distributions by acceptance--rejection (i.e., candidates are repeatedly drawn from a normal distribution until the selected parameter is positive). 
In mathematical notation, we sample $r^{(i)}\sim \mathcal{N}_+(\mu_r,\sigma_r)$ and $S^{(i)}_{\max} \sim \mathcal{N}_+(\mu_s,\sigma_s)$.
For scaling, we choose $(\mu_r,\sigma_r) = (0.2,0.02)$ and $(\mu_s,\sigma_s) = (50,5)$.
This scaling allows for interpretable size units in a variety of applications (e.g., centimeters for fish length, kilograms for mammals, and microns for cells) and for both parameters, the coefficient of variation approximates a reasonable 10\%. 
The initial size of each individual is a fraction of that individual's maximum size, $x_0^{(i)} = \eta^{(i)}S_{\max}^{(i)}$ with $\eta^{(i)} \sim \mathcal{N}_+(0.05,0.01)$ and also clipped so that $\eta^{(i)} \in [10^{-3},0.5]$ to keep growth observable.

\begin{figure}[h!]
    \centering
    \includegraphics[width=0.95\linewidth]{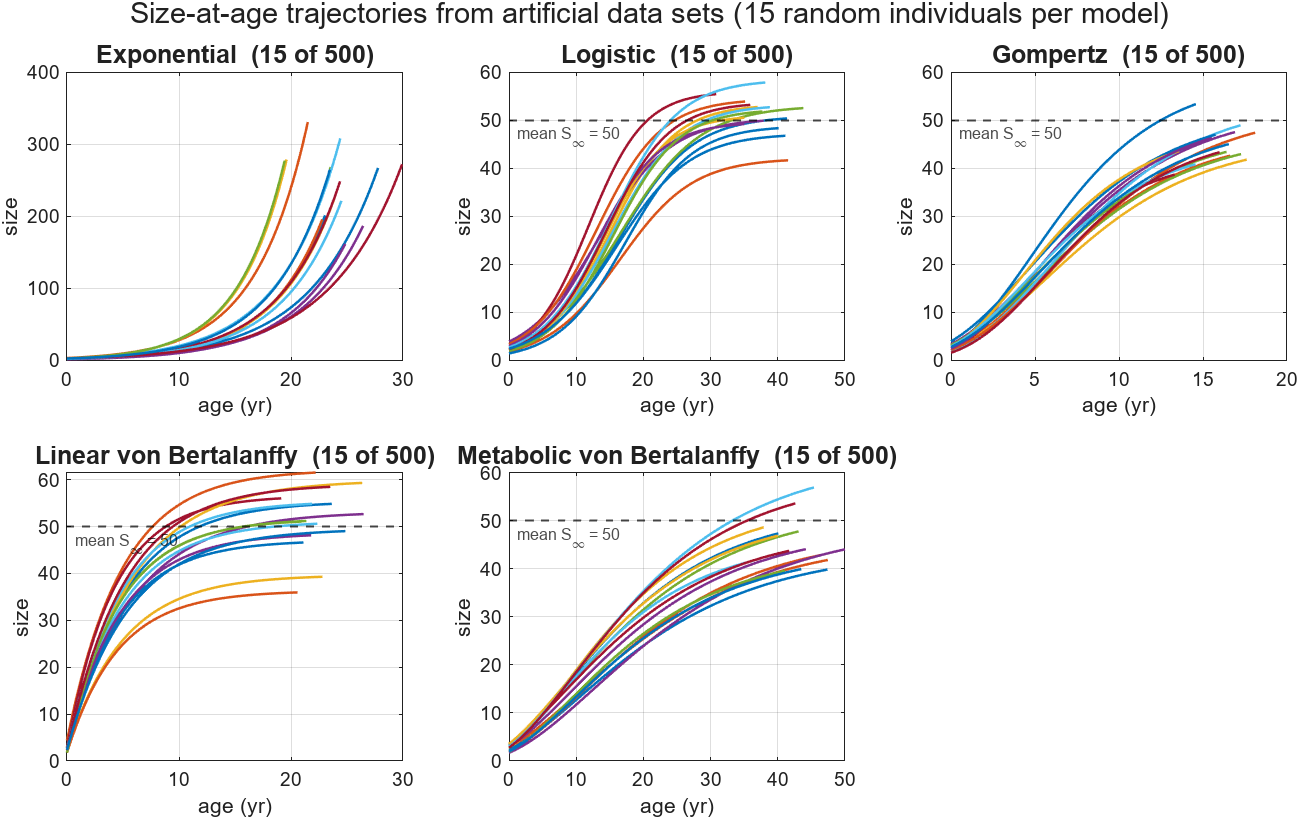}
    \caption{Size-at-age trajectory plots for each artificial data set used in the paper. }
    \label{fig:DataTraj}
\end{figure}

Each trajectory is then simulated using MATLAB's \texttt{ode45} solver with relative and absolute tolerances set to $10^{-10}$ on a uniform grid with constant sampling of $\Delta t = 0.2$.
The trajectory simulation is terminated at a ``death event" where the size crosses a particular target size $x_{T} := \min{\{100x_0, 0.9 S_{\max}\}}$.
Examples of the trajectories for each model are shown in Figure~\ref{fig:DataTraj}. 

\section{Robustness to sampling: a bootstrap experiment}
\label{asec:bootstrap}

The synthetic experiments of Section~\ref{sec:results} report point summaries computed from a single ensemble of $I=500$ individuals over many realizations of the target noise level.
Because both the individuals and their measurements are random, those summaries are effectively random variables, and a practitioner applying the method to one experimental cohort has access to exactly one draw of measurement noise. 
Therefore, it is important to quantify how much of the reported performance is attributable to the particular ensemble that happened to be observed via random sampling.

We now briefly describe the experiment.
Fix a growth law $g_\ell \in \mathcal{G}$ and a noise ratio $\sigma_{\mathrm{nr}}$,
and let $\mathcal{D} = \{({x^{(i)}}^\star, \theta^{(i)})\}_{i=1}^{I}$ denote the
noise-free synthetic ensemble generated as in Appendix~\ref{asec:ArtificialSims}.
For each of $B$ replicates $b = 1,\dots,B$ we
\begin{enumerate}
  \item draw indices $i_1^{(b)},\dots,i_n^{(b)}$ i.i.d.\ uniformly from
        $\{1,\dots,I\}$, i.e.\ resample $n = I$ individuals \emph{with
        replacement};
  \item draw an independent additive-noise realization for each resampled
        individual, $x^{(i_j)} = x^{(i_j)}_\star + \varepsilon^{(b,j)}$ with
        $\varepsilon^{(b,j)} \sim \mathcal{N}(0,\hat\sigma_{i_j}^2 I)$ and
        $\hat\sigma_{i_j} = \sigma_{\mathrm{nr}}\|x^{(i_j)}_\star\|_{\mathrm{RMS}}$;
  \item rerun the \emph{entire} model selection process of Section~\ref{sec:wendy_methods}.
\end{enumerate}
Nothing is cached between replicates, so the induced spread reflects the \emph{total} variability of the method with respect to which individuals are sampled and how they are measured. 
Individuals drawn more than once in a replicate receive independent noise realizations.

\begin{figure}[h]
    \centering
    \includegraphics[width=0.5\linewidth]{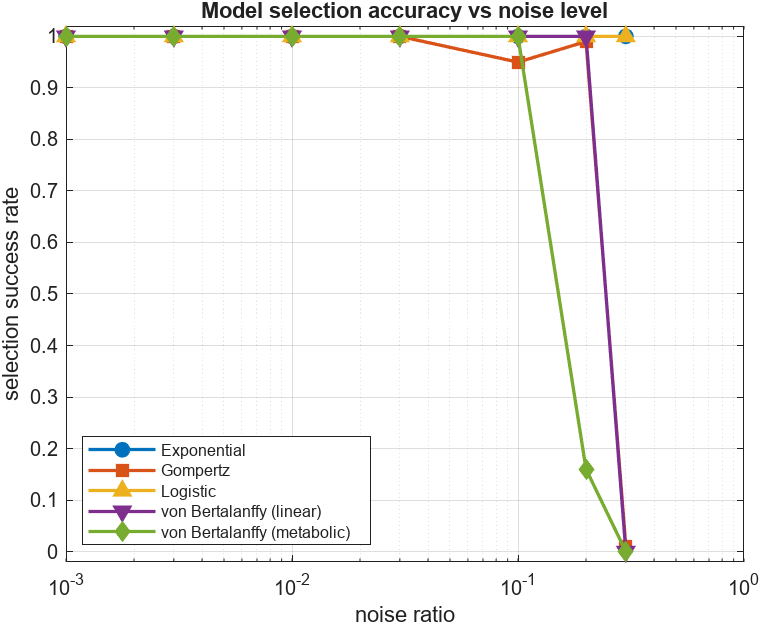}
    \caption{Selection success rate as a function of the noise ratio, $\sigma_{\mathrm{nr}}$ for $B = 100$ independent samplings of the trajectories. }
    \label{fig:Bootstrap}
\end{figure}

We test the model selection's robustness to sampling using $B = 100$ subsamples. The selection rate, interpreted as the percent of times the model was correctly selected, is reported in Fig.~\ref{fig:Bootstrap}. 
We see a similar behavior to that of Fig.~\ref{fig:sucessrate} where model selection begins to drastically fail at around $\sigma_{\mathrm{nr}} \approx 0.25$.
This suggests the model selection step (and its failure) is not strongly dependent on the given trajectories. 

\end{document}